\input harvmac
%\draftmode
\let\includefigures=\iftrue
\let\useblackboard=\iftrue
\newfam\black

%Figure Stuff
\includefigures
\message{If you do not have epsf.tex (to include figures),}
\message{change the option at the top of the tex file.}
\input epsf
\def\figin{\epsfcheck\figin}\def\figins{\epsfcheck\figins}
\def\epsfcheck{\ifx\epsfbox\UnDeFiNeD
\message{(NO epsf.tex, FIGURES WILL BE IGNORED)}
\gdef\figin##1{\vskip2in}\gdef\figins##1{\hskip.5in}% blank space instead
\else\message{(FIGURES WILL BE INCLUDED)}%
\gdef\figin##1{##1}\gdef\figinbs##1{##1}\fi}
\def\DefWarn#1{}
\def\figinsert{\goodbreak\midinsert}
\def\ifig#1#2#3{\DefWarn#1\xdef#1{fig.~\the\figno}
\writedef{#1\leftbracket fig.\noexpand~\the\figno}%
\figinsert\figin{\centerline{#3}}\medskip\centerline{\vbox{
\baselineskip12pt\advance\hsize by -1truein
\noindent\footnotefont{\bf Fig.~\the\figno:} #2}}
%\bigskip
\endinsert\global\advance\figno by1}
%%%
\else
\def\ifig#1#2#3{\xdef#1{fig.~\the\figno}
\writedef{#1\leftbracket fig.\noexpand~\the\figno}%
%\figinsert\figin{\centerline{#3}}\medskip
%\centerline{\vbox{\baselineskip12pt
%\advance\hsize by -1truein\noindent
%\footnotefont{\bf Fig.~\the\figno:} #2}}
%\bigskip\endinsert
\global\advance\figno by1} \fi

\def\id{{1 \kern-.28em {\rm l}}}

\def\K3{{\bf K3}}
\def\journal#1&#2(#3){\unskip, \sl #1\ \bf #2 \rm(19#3) }
\def\andjournal#1&#2(#3){\sl #1~\bf #2 \rm (19#3) }

\def\bar{\overline}

\def\ie{{\it i.e.}}
\def\eg{{\it e.g.}}

\def\frac#1#2{{#1\over#2}}

\def\half{\frac12}

\def\inbar{\,\vrule height1.5ex width.4pt depth0pt}
\def\IC{\relax\hbox{$\inbar\kern-.3em{\rm C}$}}
\def\IR{\relax{\rm I\kern-.18em R}}
\def\IZ{\relax{\rm I\kern-.18em Z}}

%
%%%%%%%%%%%%%%%%%%%%%%%%%%%%%%%%%%%%
%

%
\catcode`\@=11
\def\slash#1{\mathord{\mathpalette\c@ncel{#1}}}
\overfullrule=0pt

\def\EE{{\cal E}}
\def\FF{{\cal F}}

\def\MM{{\cal M}}
\def\NN{{\cal N}}
\def\OO{{\cal O}}

\def\SS{{\cal S}}

\def\underrel#1\over#2{\mathrel{\mathop{\kern\z@#1}\limits_{#2}}}

\catcode`\@=12

%%%%%%%%%%%%%%%%%%%%%%%%%%%%%%%%%%%%%%%%%%%%%%%%%%%%%%%%%%%%%%

%

\def\det{{\rm det}}

\def\det{{\rm det}}
\def\exp{{\rm exp}}

%%%%%%%%%%%%%%%%%%%%%%%%%%%%%%%%%%%%%%%%%%%%%%%%%%%%%%%%%%%%%%
% new defs:

\def\ie{{\it i.e.}}
\def\eg{{\it e.g.}}

\def\p{\partial}

%\'tHooftJZ
\lref\tHooftJZ{
  G.~'t Hooft,
  ``A Planar Diagram Theory for Strong Interactions,''
Nucl.\ Phys.\  {\bf B72}, 461 (1974).
%%CITATION = CERN-TH-1786%%
}

%\KutasovFR
\lref\KutasovFR{
  D.~Kutasov, J.~Lin, A.~Parnachev,
  ``Conformal Phase Transitions at Weak and Strong Coupling,''
[arXiv:1107.2324 [hep-th]].
%%CITATION = arXiv:1107.2324%%
}

%\MiranskyPD
\lref\MiranskyPD{
  V.~A.~Miransky and K.~Yamawaki,
  ``Conformal phase transition in gauge theories,''
  Phys.\ Rev.\  D {\bf 55}, 5051 (1997)
  [Erratum-ibid.\  D {\bf 56}, 3768 (1997)]
  [arXiv:hep-th/9611142].
  %%CITATION = PHRVA,D55,5051;%%
}

%\AntonyanVW
\lref\AntonyanVW{
  E.~Antonyan, J.~A.~Harvey, S.~Jensen, D.~Kutasov,
  ``NJL and QCD from string theory,''
[hep-th/0604017].
%%CITATION = hep-th/0604017%%
}

%\AntonyanQY
\lref\AntonyanQY{
  E.~Antonyan, J.~A.~Harvey, D.~Kutasov,
  ``The Gross-Neveu Model from String Theory,''
Nucl.\ Phys.\  {\bf B776}, 93-117 (2007).
[hep-th/0608149].
%%CITATION = hep-th/0608149%%
}

%\AntonyanPG
\lref\AntonyanPG{
  E.~Antonyan, J.~A.~Harvey, D.~Kutasov,
  ``Chiral symmetry breaking from intersecting D-branes,''
Nucl.\ Phys.\  {\bf B784}, 1-21 (2007).
[hep-th/0608177].
%%CITATION = hep-th/0608177%%
}

%\SenNF
\lref\SenNF{
  A.~Sen,
  ``Tachyon dynamics in open string theory,''
  Int.\ J.\ Mod.\ Phys.\  A {\bf 20}, 5513 (2005)
  [arXiv:hep-th/0410103].
  %%CITATION = IMPAE,A20,5513;%%
}

%\BergshoeffDQ
\lref\BergshoeffDQ{
  E.~A.~Bergshoeff, M.~de Roo, T.~C.~de Wit, E.~Eyras and S.~Panda,
  %``T duality and actions for nonBPS D-branes,''
JHEP {\bf 0005}, 009 (2000).
[hep-th/0003221].
%%CITATION = hep-th/0003221%%
}

%\KutasovDJ
\lref\KutasovDJ{
  D.~Kutasov,
  ``D-brane dynamics near NS5-branes,''
[hep-th/0405058].
%%CITATION = hep-th/0405058%%
}

%\KutasovCT
\lref\KutasovCT{
  D.~Kutasov,
  ``A geometric interpretation of the open string tachyon,''
  arXiv:hep-th/0408073.
  %%CITATION = HEP-TH/0408073;%%
}

%\SenCZ
\lref\SenCZ{
  A.~Sen,
  ``Geometric tachyon to universal open string tachyon,''
  JHEP {\bf 0705}, 035 (2007)
  [arXiv:hep-th/0703157].
  %%CITATION = JHEPA,0705,035;%%
}

%\SenAN
\lref\SenAN{
  A.~Sen,
  ``Field theory of tachyon matter,''
Mod.\ Phys.\ Lett.\  {\bf A17}, 1797-1804 (2002).
[hep-th/0204143].
%%CITATION = hep-th/0204143%%
}

%\KutasovER
\lref\KutasovER{
  D.~Kutasov, V.~Niarchos,
  ``Tachyon effective actions in open string theory,''
Nucl.\ Phys.\  {\bf B666}, 56-70 (2003).
[hep-th/0304045].
%%CITATION = hep-th/0304045%%
}

%\SenTM
\lref\SenTM{
  A.~Sen,
  ``Dirac-Born-Infeld action on the tachyon kink and vortex,''
Phys.\ Rev.\  {\bf D68}, 066008 (2003).
[hep-th/0303057].
%%CITATION = hep-th/0303057%%
}

\lref\LLL{L. Landau and E. Lifshitz, ``Quantum Mechanics,''  p. 114-117.}

%\BatellME
\lref\BatellME{
  B.~Batell, T.~Gherghetta, D.~Sword,
  ``The Soft-Wall Standard Model,''
Phys.\ Rev.\  {\bf D78}, 116011 (2008).
[arXiv:0808.3977 [hep-ph]].
%%CITATION = arXiv:0808.3977%%
}

%\MinahanTF
\lref\MinahanTF{
  J.~A.~Minahan, B.~Zwiebach,
  ``Effective tachyon dynamics in superstring theory,''
JHEP {\bf 0103}, 038 (2001).
[hep-th/0009246].
%%CITATION = hep-th/0009246%%
}

%\KutasovAQ
\lref\KutasovAQ{
  D.~Kutasov, M.~Marino and G.~W.~Moore,
  ``Remarks on tachyon condensation in superstring field theory,''
  arXiv:hep-th/0010108.
  %%CITATION = HEP-TH/0010108;%%
}

%\WittenZW
\lref\WittenZW{
  E.~Witten,
  ``Anti-de Sitter space, thermal phase transition, and confinement in gauge theories,''
Adv.\ Theor.\ Math.\ Phys.\  {\bf 2}, 505-532 (1998).
[hep-th/9803131].
%%CITATION = hep-th/9803131%%
}

%\SakaiCN
\lref\SakaiCN{
  T.~Sakai, S.~Sugimoto,
  ``Low energy hadron physics in holographic QCD,''
Prog.\ Theor.\ Phys.\  {\bf 113}, 843-882 (2005).
[arXiv:hep-th/0412141 [hep-th]].
%%CITATION = IU-MSTP-63%%
}

%\ErlichQH
\lref\ErlichQH{
  J.~Erlich, E.~Katz, D.~T.~Son, M.~A.~Stephanov,
  "QCD and a holographic model of hadrons,''
Phys.\ Rev.\ Lett.\  {\bf 95}, 261602 (2005).
[hep-ph/0501128].
%%CITATION = hep-ph/0501128%%
}

%\KaplanKR
\lref\KaplanKR{
  D.~B.~Kaplan, J.~-W.~Lee, D.~T.~Son, M.~A.~Stephanov,
  ``Conformality Lost,''
Phys.\ Rev.\  {\bf D80}, 125005 (2009).
[arXiv:0905.4752 [hep-th]].
%%CITATION = arXiv:0905.4752%%
}

%\DaRoldZS
\lref\DaRoldZS{
  L.~Da Rold and A.~Pomarol,
  ``Chiral symmetry breaking from five dimensional spaces,''
  Nucl.\ Phys.\  B {\bf 721}, 79 (2005)
  [arXiv:hep-ph/0501218].
  %%CITATION = NUPHA,B721,79;%%
}

%\KarchPV
\lref\KarchPV{
  A.~Karch, E.~Katz, D.~T.~Son, M.~A.~Stephanov,
  ``Linear confinement and AdS/QCD,''
Phys.\ Rev.\  {\bf D74}, 015005 (2006).
[hep-ph/0602229].
%%CITATION = hep-ph/0602229%%
}

%\WittenZW
\lref\WittenZW{
  E.~Witten,
  ``Anti-de Sitter space, thermal phase transition, and confinement in  gauge
  theories,''
  Adv.\ Theor.\ Math.\ Phys.\  {\bf 2}, 505 (1998)
  [arXiv:hep-th/9803131].
  %%CITATION = 00203,2,505;%%
}

%\KlebanovHB
\lref\KlebanovHB{
  I.~R.~Klebanov, M.~J.~Strassler,
  ``Supergravity and a confining gauge theory: Duality cascades and chi SB resolution of naked singularities,''
JHEP {\bf 0008}, 052 (2000).
[arXiv:hep-th/0007191 [hep-th]].
%%CITATION = IASSNS-HEP-00-56%%
}

%\HerzogRA
\lref\HerzogRA{
  C.~P.~Herzog,
  ``A holographic prediction of the deconfinement temperature,''
  Phys.\ Rev.\ Lett.\  {\bf 98}, 091601 (2007)
  [arXiv:hep-th/0608151].
  %%CITATION = PRLTA,98,091601;%%
}

%\MateosNU
\lref\MateosNU{
  D.~Mateos, R.~C.~Myers and R.~M.~Thomson,
  ``Holographic phase transitions with fundamental matter,''
  Phys.\ Rev.\ Lett.\  {\bf 97}, 091601 (2006)
  [arXiv:hep-th/0605046].
  %%CITATION = PRLTA,97,091601;%%
}

%\ParnachevDN
\lref\ParnachevDN{
  A.~Parnachev, D.~A.~Sahakyan,
  ``Chiral Phase Transition from String Theory,''
Phys.\ Rev.\ Lett.\  {\bf 97}, 111601 (2006).
[hep-th/0604173].
%%CITATION = hep-th/0604173%%
}

%\AharonyDA
\lref\AharonyDA{
  O.~Aharony, J.~Sonnenschein, S.~Yankielowicz,
  ``A Holographic model of deconfinement and chiral symmetry restoration,''
Annals Phys.\  {\bf 322}, 1420-1443 (2007).
[hep-th/0604161].
%%CITATION = hep-th/0604161%%
}

%\KruczenskiUQ
\lref\KruczenskiUQ{
  M.~Kruczenski, D.~Mateos, R.~C.~Myers and D.~J.~Winters,
  ``Towards a holographic dual of large-N(c) QCD,''
  JHEP {\bf 0405}, 041 (2004)
  [arXiv:hep-th/0311270].
  %%CITATION = JHEPA,0405,041;%%
}

%\GursoyJH
\lref\GursoyJH{
  U.~Gursoy,
  ``Continuous Hawking-Page transitions in Einstein-scalar gravity,''
  JHEP {\bf 1101}, 086 (2011)
  [arXiv:1007.0500 [hep-th]].
  %%CITATION = JHEPA,1101,086;%%
}

%\HartnollKX
\lref\HartnollKX{
  S.~A.~Hartnoll, C.~P.~Herzog, G.~T.~Horowitz,
  ``Holographic Superconductors,''
JHEP {\bf 0812}, 015 (2008).
[arXiv:0810.1563 [hep-th]].
%%CITATION = arXiv:0810.1563%%
}

%\CaseroAE
\lref\CaseroAE{
  R.~Casero, E.~Kiritsis and A.~Paredes,
  ``Chiral symmetry breaking as open string tachyon condensation,''
  Nucl.\ Phys.\  B {\bf 787}, 98 (2007)
  [arXiv:hep-th/0702155].
  %%CITATION = NUPHA,B787,98;%%
}

%\BatellZM
\lref\BatellZM{
  B.~Batell and T.~Gherghetta,
  ``Dynamical Soft-Wall AdS/QCD,''
  Phys.\ Rev.\  D {\bf 78}, 026002 (2008)
  [arXiv:0801.4383 [hep-ph]].
  %%CITATION = PHRVA,D78,026002;%%
}

%\SemenoffJF
\lref\SemenoffJF{
  G.~W.~Semenoff,
  ``Chiral Symmetry Breaking in Graphene,''
[arXiv:1108.2945 [hep-th]].
%%CITATION = arXiv:1108.2945%%
}

%\HillAP
\lref\HillAP{
  C.~T.~Hill and E.~H.~Simmons,
  ``Strong dynamics and electroweak symmetry breaking,''
  Phys.\ Rept.\  {\bf 381}, 235 (2003)
  [Erratum-ibid.\  {\bf 390}, 553 (2004)]
  [arXiv:hep-ph/0203079].
  %%CITATION = PRPLC,381,235;%%
}

%\CohenSQ
\lref\CohenSQ{
  A.~G.~Cohen, H.~Georgi,
  ``Walking Beyond The Rainbow,''
Nucl.\ Phys.\  {\bf B314}, 7 (1989).
%%CITATION = HUTP-88/A007%%
}

%\KarchEG
\lref\KarchEG{
  A.~Karch, E.~Katz, D.~T.~Son, M.~A.~Stephanov,
  ``On the sign of the dilaton in the soft wall models,''
JHEP {\bf 1104}, 066 (2011).
[arXiv:1012.4813 [hep-ph]].
%%CITATION = arXiv:1012.4813%%
}

%\HartnollFN
\lref\HartnollFN{
  S.~A.~Hartnoll,
  ``Horizons, holography and condensed matter,''
[arXiv:1106.4324 [hep-th]].
%%CITATION = arXiv:1106.4324%%
}

%\HerzogXV
\lref\HerzogXV{
  C.~P.~Herzog,
  ``Lectures on Holographic Superfluidity and Superconductivity,''
J.\ Phys.\ AA\ {\bf 42}, 343001  (2009).
[arXiv:0904.1975 [hep-th]].
%%CITATION = arXiv:0904.1975%%
}

%\McGreevyXE
\lref\McGreevyXE{
  J.~McGreevy,
  ``Holographic duality with a view toward many-body physics,''
Adv.\ High Energy Phys.\ \ {\bf 2010}, 723105  (2010).
[arXiv:0909.0518 [hep-th]].
%%CITATION = arXiv:0909.0518%%
}

%\HorowitzGK
\lref\HorowitzGK{
  G.~T.~Horowitz,
  ``Introduction to Holographic Superconductors,''
[arXiv:1002.1722 [hep-th]].
%%CITATION = arXiv:1002.1722%%
}

%\SachdevWG
\lref\SachdevWG{
  S.~Sachdev,
  ``What can gauge-gravity duality teach us about condensed matter physics?,''
  arXiv:1108.1197 [cond-mat.str-el].
  %%CITATION = ARXIV:1108.1197;%%
}

%\DymarskyNC
\lref\DymarskyNC{
  A.~Dymarsky, I.~R.~Klebanov and R.~Roiban,
  ``Perturbative gauge theory and closed string tachyons,''
JHEP\ {\bf 0511}, 038  (2005).
[hep-th/0509132].
%%CITATION = hep-th/0509132%%
}

%\DymarskyUH
\lref\DymarskyUH{
  A.~Dymarsky, I.~R.~Klebanov and R.~Roiban,
  ``Perturbative search for fixed lines in large N gauge theories,''
JHEP\ {\bf 0508}, 011  (2005).
[hep-th/0505099].
%%CITATION = hep-th/0505099%%
}

%\AdamsJB
\lref\AdamsJB{
  A.~Adams and E.~Silverstein,
  ``Closed string tachyons, AdS / CFT, and large N QCD,''
Phys.\ Rev.\ D\ {\bf 64}, 086001  (2001).
[hep-th/0103220].
%%CITATION = hep-th/0103220%%
}

%\PomoniDE
\lref\PomoniDE{
  E.~Pomoni and L.~Rastelli,
  ``Large N Field Theory and AdS Tachyons,''
JHEP\ {\bf 0904}, 020  (2009).
[arXiv:0805.2261 [hep-th]].
%%CITATION = arXiv:0805.2261%%
}

%\NunezWI
\lref\NunezWI{
  C.~Nunez, I.~Papadimitriou and M.~Piai,
  ``Walking Dynamics from String Duals,''
Int.\ J.\ Mod.\ Phys.\ A\ {\bf 25}, 2837  (2010).
[arXiv:0812.3655 [hep-th]].
%%CITATION = arXiv:0812.3655%%
}

%\ElanderPK
\lref\ElanderPK{
  D.~Elander, C.~Nunez and M.~Piai,
  ``A Light scalar from walking solutions in gauge-string duality,''
Phys.\ Lett.\ B\ {\bf 686}, 64  (2010).
[arXiv:0908.2808 [hep-th]].
%%CITATION = arXiv:0908.2808%%
}

%\NunezDA
\lref\NunezDA{
  C.~Nunez, M.~Piai and A.~Rago,
  ``Wilson Loops in string duals of Walking and Flavored Systems,''
Phys.\ Rev.\ D\ {\bf 81}, 086001  (2010).
[arXiv:0909.0748 [hep-th]].
%%CITATION = arXiv:0909.0748%%
}

%\JarvinenFE
\lref\JarvinenFE{
  M.~Jarvinen and F.~Sannino,
  ``Holographic Conformal Window - A Bottom Up Approach,''
JHEP\ {\bf 1005}, 041  (2010).
[arXiv:0911.2462 [hep-ph]].
%%CITATION = arXiv:0911.2462%%
}

%\HabaHU
\lref\HabaHU{
  K.~Haba, S.~Matsuzaki and K.~Yamawaki,
  ``Holographic Techni-dilaton,''
Phys.\ Rev.\ D\ {\bf 82}, 055007  (2010).
[arXiv:1006.2526 [hep-ph]].
%%CITATION = arXiv:1006.2526%%
}

%\PremKumarAS
\lref\PremKumarAS{
  S.~Prem Kumar, D.~Mateos, A.~Paredes and M.~Piai,
  ``Towards holographic walking from N=4 super Yang-Mills,''
JHEP\ {\bf 1105}, 008  (2011).
[arXiv:1012.4678 [hep-th]].
%%CITATION = arXiv:1012.4678%%
}

%\AnguelovaBC
\lref\AnguelovaBC{
  L.~Anguelova, P.~Suranyi and L.~C.~R.~Wijewardhana,
  ``Holographic Walking Technicolor from D-branes,''
Nucl.\ Phys.\ B\ {\bf 852}, 39  (2011).
[arXiv:1105.4185 [hep-th]].
%%CITATION = arXiv:1105.4185%%
}

%\JK
\lref\JK{
  M.~Jarvinen and E.~Kiritsis,
  ``Holographic Models for QCD in the Veneziano Limit,''
[arXiv:1112.1261 [hep-ph]].
%%CITATION = arXiv:1112.1261%%
}

%\ShifmanXN
\lref\ShifmanXN{
  M.~Shifman and A.~Vainshtein,
  ``Highly Excited Mesons, Linear Regge Trajectories and the Pattern of the Chiral Symmetry Realization,''
Phys.\ Rev.\ D\ {\bf 77}, 034002  (2008).
[arXiv:0710.0863 [hep-ph]].
%%CITATION = arXiv:0710.0863%%
}

%\CampbellIW
\lref\CampbellIW{
  B.~A.~Campbell, J.~Ellis and K.~A.~Olive,
  ``Phenomenology and Cosmology of an Electroweak Pseudo-Dilaton and Electroweak Baryons,''
[arXiv:1111.4495 [hep-ph]].
%%CITATION = arXiv:1111.4495%%
}

%\JensenGA
\lref\JensenGA{
  K.~Jensen, A.~Karch, D.~T.~Son and E.~G.~Thompson,
  ``Holographic Berezinskii-Kosterlitz-Thouless Transitions,''
Phys.\ Rev.\ Lett.\  {\bf 105}, 041601 (2010).
[arXiv:1002.3159 [hep-th]].
%%CITATION = arXiv:1002.3159%%
}

%\IqbalEH
\lref\IqbalEH{
  N.~Iqbal, H.~Liu, M.~Mezei and Q.~Si,
  ``Quantum phase transitions in holographic models of magnetism and
  superconductors,''
  Phys.\ Rev.\  D {\bf 82}, 045002 (2010)
  [arXiv:1003.0010 [hep-th]].
  %%CITATION = PHRVA,D82,045002;%%
}

%\EvansHI
\lref\EvansHI{
  N.~Evans, A.~Gebauer, K.~Y.~Kim and M.~Magou,
  ``Phase diagram of the D3/D5 system in a magnetic field and a BKT
  transition,''
  Phys.\ Lett.\  B {\bf 698}, 91 (2011)
  [arXiv:1003.2694 [hep-th]].
  %%CITATION = PHLTA,B698,91;%%
}

%\PalGJ
\lref\PalGJ{
  S.~S.~Pal,
  ``Quantum phase transition in a Dp-Dq system,''
  Phys.\ Rev.\  D {\bf 82}, 086013 (2010)
  [arXiv:1006.2444 [hep-th]].
  %%CITATION = PHRVA,D82,086013;%%
}

%\JensenVX
\lref\JensenVX{
  K.~Jensen,
  ``More Holographic Berezinskii-Kosterlitz-Thouless Transitions,''
  Phys.\ Rev.\  D {\bf 82}, 046005 (2010)
  [arXiv:1006.3066 [hep-th]].
  %%CITATION = PHRVA,D82,046005;%%
}

%\JensenAF
\lref\JensenAF{
  K.~Jensen,
  ``Semi-Holographic Quantum Criticality,''
Phys.\ Rev.\ Lett.\  {\bf 107}, 231601 (2011).
[arXiv:1108.0421 [hep-th]].
%%CITATION = arXiv:1108.0421%%
}

%\ReyZZ
\lref\ReyZZ{
  S.~-J.~Rey,
  ``String theory on thin semiconductors: Holographic realization of Fermi points and surfaces,''
%PTPSA,177,128-142.\ 2009 {\bf 177}, 128 (2009).
[arXiv:0911.5295 [hep-th]].
%%CITATION = arXiv:0911.5295%%
}

%\IqbalAJ
\lref\IqbalAJ{
  N.~Iqbal, H.~Liu and M.~Mezei,
  ``Quantum phase transitions in semi-local quantum liquids,''
[arXiv:1108.0425 [hep-th]].
%%CITATION = arXiv:1108.0425%%
}

%\IatrakisJB
\lref\IatrakisJB{
  I.~Iatrakis, E.~Kiritsis and A.~Paredes,
  ``An AdS/QCD model from tachyon condensation: II,''
JHEP {\bf 1011}, 123 (2010).
[arXiv:1010.1364 [hep-ph]].
%%CITATION = arXiv:1010.1364%%
}

%\BigazziMD
\lref\BigazziMD{
  F.~Bigazzi, R.~Casero, A.~L.~Cotrone, E.~Kiritsis and A.~Paredes,
  %``Non-critical holography and four-dimensional CFT's with fundamentals,''
JHEP {\bf 0510}, 012 (2005).
[hep-th/0505140].
%%CITATION = hep-th/0505140%%
}

%%%%%%%%%%%%%%%%%%%%%%%%%%%%%%%%%%%%%%%%%%%%%%%%%%%
\Title{}
{\vbox{\centerline{Holographic Walking from Tachyon DBI}
%\bigskip
%\centerline{}
}}
\bigskip

\centerline{\it  David Kutasov$^1$, Jennifer Lin$^1$ and Andrei Parnachev$^{2}$}
\bigskip
\smallskip
\centerline{${}^{1}$EFI and Department of Physics, University of
Chicago} \centerline{5640 S. Ellis Av., Chicago, IL 60637, USA }
\smallskip
\centerline{${}^{2}$Institute Lorentz for Theoretical Physics, Leiden University} 
\centerline{P.O. Box 9506, Leiden 2300RA, The Netherlands}
\smallskip

\vglue .3cm

\bigskip

\let\includefigures=\iftrue
\bigskip
\noindent
We use holography to study Conformal Phase Transitions, which are believed to be realized in four dimensional QCD and play an important role in walking technicolor models of electroweak symmetry breaking. At strong coupling they can be modeled by the non-linear dynamics of a tachyonic scalar field with mass close to the Breitenlohner-Freedman bound in anti de Sitter spacetime. Taking the action for this field to have  a Tachyon-Dirac-Born-Infeld form gives rise to models that resemble hard and soft wall AdS/QCD, with a dynamically generated wall. For hard wall models, the highly excited spectrum has the KK form $m_n\sim n$; in the soft wall case we exhibit potentials with $m_n\sim n^\alpha$, $0<\alpha\le 1/2$. We investigate the finite temperature phase structure and find first or second order symmetry restoration transitions, depending on the behavior of the potential near the origin of field space.

\bigskip

\Date{}

\newsec{Introduction}

Consider $SU(N)$ gauge theory coupled to $F$ massless Dirac fermions in the fundamental representation of the gauge group.\foot{It is convenient for our purposes to study this theory in the limit $F,N\gg 1$, so that the parameter $F/N$ can be treated as continuous. Much of what we say can be extended to finite $F,N$.} For $F\ge 11N/2$, this theory is free at long distances, and its low energy dynamics can be studied using the techniques of perturbative field theory. For $F<11N/2$, the theory becomes interacting in the infrared where it is said to be in a non-abelian Coulomb phase. The crossover between the free ultraviolet behavior and the interacting infrared conformal field theory occurs at a scale $\Lambda_{QCD}$. As $F$ decreases, the infrared coupling increases; eventually, perturbation theory fails at scales of order $\Lambda_{QCD}$ and below.   

On the other hand, when $F\ll N$ the theory is believed to confine and break the chiral symmetry $SU(F)_L\times SU(F)_R\to SU(F)_{\rm diag}$. The 't Hooft large $N$ analysis of perturbation theory \tHooftJZ\ provides a nice picture of what happens in this regime. To leading order in $F/N$, the only fields that run in loops are the adjoint degrees of freedom (the gauge fields). The spectrum includes $F^2-1$ massless Nambu-Goldstone bosons (``pions''), and a discrete spectrum of massive glueballs and mesons with masses of order $\Lambda_{QCD}$. 

\ifig\loc{Schematic phase diagram of QCD as a function of the number of flavors $F$. $\Lambda_{QCD}$ is the scale above which the theory becomes free; $\mu$ is the meson mass scale.}
{\epsfxsize2.7in\epsfbox{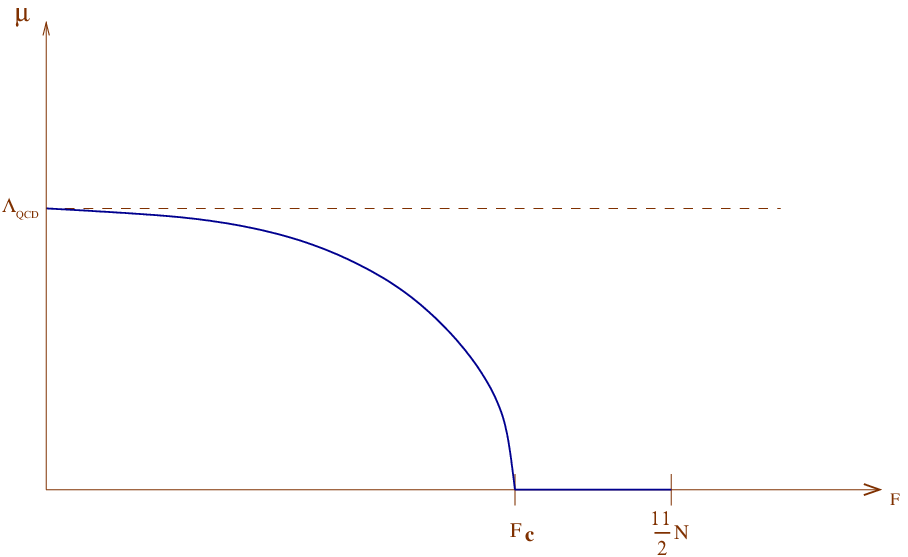}}

It is natural to ask how the above two pictures are connected as one varies the number of flavors $F$. Since the realization  of chiral symmetry is different in the two regimes, there must be a phase transition at a finite value of $F/N$. This transition is believed to be continuous; the order parameter, which can be taken to be the typical meson mass $\mu$, is expected to behave as a function of $F$ in the way depicted  in figure 1. According to this picture, the non-abelian Coulomb phase extends down to a critical number of flavors, $F_c$, which is not known precisely but is believed to be $F_c\simeq 4N$. When $F$ decreases past $F_c$,  $\mu$ becomes non-zero, but near the transition it is much smaller than the scale $\Lambda_{QCD}$ above which the theory becomes free. As $F$ decreases further, $\mu$ increases and eventually approaches $\Lambda_{QCD}$.  

The transition at $F=F_c$ is not well understood. A widely discussed scenario (see \eg\ \CohenSQ\ and references therein) is that it is driven by the gauge invariant fermion bilinear operator $\bar\psi\psi$. The UV scaling dimension of this operator is three, but in the non-abelian Coulomb phase its IR scaling dimension is lowered by  gauge interactions. It is believed that as $F\to F_c$, the IR dimension approaches two, and when $F<F_c$ it becomes complex. This behavior signals an instability of the Coulomb phase due to a non-vanishing $\beta$-function for the ``double trace'' operator $({\rm Tr}\bar\psi\psi)^2$, which leads to the appearance of the small mass gap $\mu$ in figure 1. Near $F_c$, the gap scales like $\mu\sim \Lambda_{QCD}\exp(-a/\sqrt{F-F_c})$, which is usually referred to as Miransky scaling. The corresponding phase transition is known as a Conformal Phase Transition (CPT)  \MiranskyPD.

Understanding the behavior of the theory in the vicinity of the phase transition at $F=F_c$ is an interesting open problem, which may also be important for applications. In particular, a version of technicolor known as ``walking technicolor'' relies on it  to generate electroweak symmetry breaking \HillAP, and similar dynamics may have applications to condensed matter systems such as graphene (see \eg\ \SemenoffJF\ for a recent discussion).   

In thinking about the physics near the transition, it is useful to divide the RG flow of QCD into two stages. The first involves the  flow from an asymptotically free field theory at energies well above $\Lambda_{QCD}$ to a strongly coupled CFT well below that scale. The only part of this flow which is of interest for studying the transition is its endpoint, the non-trivial infrared CFT which serves as the arena for the transition. Therefore, for our purposes we can treat $\Lambda_{QCD}$ as a UV cutoff, and focus on the dynamics well below it. For $F\ge F_c$, the infrared CFT is stable under RG, while for $F<F_c$ it dynamically generates the scale $\mu$. The problem one is faced with is to understand the mechanism of mass generation. 

As stressed in \KaplanKR, the behavior exhibited by this example occurs much more generally, whenever two fixed points of the RG approach each other, merge and move off into the complex plane. In general (at large $N$), the transition is driven by an operator $\OO$ whose dimension approaches $d/2$ near the critical point. In QCD, $d=4$ and $\OO={\rm Tr}\bar\psi\psi$; examples with other values of $d$, which involve systems of fermions localized on defects interacting with gauge fields in the bulk were recently discussed in \KutasovFR.

The fact that the transition is continuous makes it natural to expect that it can be described  by studying the dynamics of the order parameter $\OO$. A conventional Landau-Ginzburg description is inappropriate here, since the transition is essentially an infinite order one (as reflected \eg\ in the Miransky scaling). A further difficulty is that the dynamics is typically strongly coupled in the vicinity of the transition. It is natural to ask whether holography can be used to shed light on this problem. In this paper, we will take some steps towards this goal. Other recent discussions of holographic walking include \refs{\NunezWI\ElanderPK\NunezDA\JarvinenFE\HabaHU\PremKumarAS\AnguelovaBC-\JK}.

The starting point of our discussion is the assumption that the $d$ dimensional CFT that serves as the arena in which the phase transition occurs has an $AdS_{d+1}$ dual. The order parameter $\OO$ corresponds in the bulk description to a  scalar field $T$ whose mass approaches the Breitenlohner-Freedman (BF) bound. Thus, to describe the physics associated with the scale $\mu$ in figure 1, we need to study the dynamics of $T$ in $AdS_{d+1}$. 

Using the standard 't Hooft map \tHooftJZ, the field $T$ can often be thought of as an open string tachyon.\foot{This is certainly true in probe brane systems, such as those studied in \KutasovFR, and formally is also true in QCD, although for $F\sim N$ the distinction between open and closed strings is blurred.} The dynamics of such tachyons has been extensively studied in the past and a lot is known about it (see \eg\ \SenNF\ for a review). One of the interesting outcomes of these studies is the realization that the dynamics of open string tachyons is well described qualitatively, and in some cases quantitatively, by an effective action called the Tachyon-Dirac-Born-Infeld (TDBI) action  \refs{\SenAN\SenTM-\KutasovER}. Thus, it is natural to model the phase transition using such an action in $AdS_{d+1}$.  

The TDBI action depends on the choice of a potential $V(T)$ for the tachyon. The quadratic term in the potential gives the mass of the tachyon, which we will take to be close to the BF bound. As we will see, the low energy dynamics depends on the full potential, and one of our main goals will be to explore this dependence. This can be viewed as a bottom-up approach, analogous to the one taken in AdS/QCD. We will see that when the mass of the tachyon is slightly above the BF bound, the theory is in a conformal phase, but when the mass is slightly below the bound, it breaks conformal symmetry and generates a mass gap. This gap can be made parametrically small relative to the UV cutoff $\Lambda$, which plays the role of $\Lambda_{QCD}$. Depending on the choice of potential, we find different spectra of mesons, including some that agree with what is expected in QCD in the 't Hooft limit. 

We will also explore the thermodynamics of TDBI models and find that they exhibit first or second order phase transitions at a finite temperature, depending on the small $T$ behavior of the potential $V(T)$. This differs from most other holographic models, which exhibit strongly first order phase transitions.  We will discuss the origin of the differences between our model and other models of holographic QCD in section 7.

The plan of the paper is as follows. In section 2 we revisit a system that we studied in  \KutasovFR, $N=4$ SYM coupled to fermions in the fundamental representation of the gauge group $U(N)$ localized on a $d-1$ dimensional defect. There, we showed that this system exhibits a conformal phase transition, and studied it using holography. In section 2 we show that the discussion of \KutasovFR\ can be rephrased in terms of a TDBI action for a real tachyon field $T$, and we compute the potential $V(T)$. This provides support  for the idea that one can use the TDBI action to describe CPTs such as the one expected in QCD. 

In section 3 we discuss the TDBI approach to general CPTs.  We describe the vacuum solution for different tachyon potentials, and in particular show that the TDBI model gives an interesting dynamical  realization of soft \ErlichQH\ and hard \KarchPV\ wall models of AdS/QCD. We also derive the Lagrangian for small fluctuations around the vacuum, which correspond to scalar and vector mesons.

In sections 4 and 5 we discuss the hard and soft wall models, respectively. We show that hard wall models always have KK-type spectra of excitations $m_n\sim n$ at large excitation number $n$, in agreement with the analysis of \KutasovFR\ which involved a particular hard wall model. On the other hand, soft wall models can give other types of asymptotic spectra. We discuss explicitly a class of potentials which gives the asymptotic spectrum $m_n\sim  n^\alpha$ with arbitrary $0<\alpha\le1/2$.  We also discuss the question of restoration of chiral symmetry in the asymptotic spectrum of (axial) vectors and (pseudo) scalars. 

In section 6 we study the thermodynamics of TDBI models. We show that they undergo a first or second order phase transition at a finite temperature, depending on the form of the potential near the origin of field space. The structure we find is rather different than in other holographic models of confinement. In section 7 we discuss our results and comment on possible extensions. Some technical details of the meson spectrum in the axial sector, and a numerical check of the finite temperature analysis, are relegated to the appendices.

\newsec{Defect fermions coupled to $N=4$ SYM}

In \KutasovFR\ we showed that the system of $\NN=4$ SYM coupled to fermions localized on a $d-1$ dimensional defect exhibits a transition similar to that expected in QCD. While in QCD this transition occurs as a function of the number of colors and flavors, in the system studied in \KutasovFR\ the transition happens for any number of flavors, and is driven by the 't Hooft coupling $\lambda$ of $\NN=4$ SYM. As the coupling increases, the scaling dimension of the fermion bilinear $\bar\psi\psi$ decreases. When the coupling reaches a critical value $\lambda_c$, $\Delta(\bar\psi\psi)$ approaches $d/2$, and the system undergoes a transition from a conformal phase to one in which the fermions are massive. This is an example of  what we referred to in the introduction as a conformal phase transition (CPT). 

The defect theory can be embedded in string theory as the low energy theory on $N$ $D3$-branes and one or more $Dp$-branes, which intersect the $D3$-branes on a $d$ dimensional spacetime and are extended in $n$ additional directions transverse to the threebranes. Strings stretched between the $D3$ and $Dp$-branes give rise to fermions in the fundamental representation of $U(N)$. At weak coupling, the dynamics is described by a Lagrangian which couples the fermions to the $D3$-brane gauge field and $6-n=d-p+5$ adjoint scalars. At strong coupling, one can replace the $D3$-branes by their near-horizon geometry $AdS_5\times S^5$, and study the dynamics of the $Dp$-branes as probes in this geometry. 

In \KutasovFR\ we pointed out that one can study the defect system as a function of the parameters $d$, $n$, formally viewing them as continuous variables, in the spirit of the $\epsilon$ expansion. By varying these parameters, one can change the critical coupling $\lambda_c$.  For fixed $n$ and $d$ slightly above two, $\lambda_c\ll 1$ and the transition can be studied using perturbative gauge theory. For fixed $d>2$ and $n$ slightly above $1+d^2/4$, $\lambda_c$ is large, and the transition can be studied using holography. At strong coupling, the transition is between a phase in which the probe brane wraps an $AdS_{d+1}\times S^{n-1}$ inside  $AdS_5\times S^5$, and one in which the  probe brane is deformed in the IR. In \KutasovFR\ we used the gravitational description to study aspects of the physics associated with the transition. 

The purpose of this paper is to generalize the discussion of \KutasovFR\ to a larger class of systems. The eventual hope is to construct a holographic description of QCD in the vicinity of the transition from non-abelian Coulomb to confining infrared behavior. As a step in this direction, we will present a class of bottom up models whose properties are in qualitative agreement with expectations. As we will see, our construction makes contact with both the hard and soft wall models of AdS/QCD \refs{\ErlichQH,\KarchPV}, although the way they arise here is rather different than in the literature. 

We start with a brief review of some aspects of \KutasovFR. The near-horizon geometry of the $D3$-branes, $AdS_5 \times S^5$, is described by the metric
\eqn\ads{ ds^2 = \left(\frac r L\right)^2dx_\mu dx^\mu + \left(\frac L r\right)^2dr^2 + L^2d\Omega_5^2,}
where $x^\mu$, $\mu = 0,1,2,3$, are directions along the $D3$-branes, and $r$ and $\Omega_5$ are spherical coordinates on the transverse $\IR^6$. $L$ is the radius of curvature of $AdS_5$ and the size of the $S^5$.

Into this background, we place a $Dp$-brane that wraps $d$ of the four $x^\mu$ and an $\IR^n$ subspace of the transverse $\IR^6$.  Using spherical coordinates $(\rho,\Omega_{n-1})$ on $\IR^n$, we can write the metric \ads\ as
\eqn\oldmetric {ds^2 = \left(\frac r L\right)^2dx_\mu dx^\mu + \left(\frac L r\right)^2(d\rho^2 + \rho^2d\Omega_{n-1}^2 + d\phi_i^2)~. }
The $Dp$-brane wraps an  $AdS_{d+1} \times S^{n-1}$ parametrized by  $x^a$, $a=0,1,2,\cdots, d-1$, as well as $\rho$ and $\Omega_{n-1}$. The coordinates $\phi_i$, $i=1,\cdots, 6-n$, parametrize directions transverse to the probe brane. The radial coordinate $r$ is given by 
\eqn\rrrho{r^2=\rho^2+\sum_i\phi_i^2.}
Excitations of the probe brane correspond to fluctuations of the scalar fields $\phi_i$, the worldvolume gauge field, and fermions. Their dynamics, which is governed by the DBI action, was studied  in \KutasovFR.  

It is useful to note that the action of the probe brane must be $SO(d,2)$ invariant. The reason is that the action can be thought of as describing small fluctuations around the $AdS_{d+1}\times S^{n-1}$ configuration $\phi_i=0$. Although this configuration is not always dynamically stable, this does not matter for the symmetry structure. $SO(d,2)$ is the subgroup of the $SO(4,2)$ isometry of the metric  \ads, which acts on the coordinates $(r, x^a)$. 

In the DBI action derived from the metric \oldmetric, which was discussed in \KutasovFR, the above $SO(d,2)$ symmetry is hidden for the following reason. While the symmetry acts naturally on the radial coordinate $r$, we have split it here as in \rrrho, with $\rho$ a worldvolume coordinate and $\phi_i$ fields on the brane. It is thus clear how one needs to modify the analysis of \KutasovFR\ to make the symmetry manifest.\foot{We thank E. Martinec for a useful discussion of this issue.} 
The idea is to take the probe $Dp$-brane to wrap the $d$ dimensions labeled by $x^a$, the radial direction $r$, and an $S^{n-1}$ of maximal radius in the $S^5$. The $6-n$ directions on $S^5$ transverse to the $S^{n-1}$ give rise to scalar fields on the worldvolume. These scalar fields are by construction compact, since they correspond to angular variables on the sphere. Their action is manifestly invariant under $SO(d,2)$.

To make this more explicit, consider an excitation of one of the scalar fields $\theta = \theta(x^M),$ where  $(x^M) = (x^a, r)$ parametrize the $AdS_{d+1}$. The metric on the $S^5$ can be written as
\eqn\metsphere{d\Omega_5^2=L^2d\theta^2+(L\cos\theta)^2d\Omega_{n-1}^2+\cdots} 
where the ellipsis denotes terms associated with other directions on the sphere, which we do not excite. The induced metric on the $Dp$-brane now takes the form
\eqn\metrictwo{ ds^2 = \left(g_{MN}+L^2 \partial_M\theta \partial_N\theta\right) dx^Mdx^N + (L\cos\theta)^2d\Omega_{n-1}^2}
where $g_{MN}$ is the $AdS_{d+1}$ metric (\ads\ restricted to $x^M$). 
The DBI action that follows from \metrictwo\ is\foot{We omitted an overall multiplicative factor.}
\eqn\dbitwo{\SS =- \int d^{d+1}x \sqrt{-g} (\cos\theta)^{n-1}\sqrt{1 + L^2g^{MN}\partial_M\theta\partial_N\theta}. }
This action is manifestly covariant in $AdS_{d+1}$ and thus is $SO(d,2)$ symmetric. From the point of view 
of the parametrization employed in \KutasovFR, the excitation $\theta(x^M)$ corresponds to one of the $6-n$ scalar fields, say $\phi=\phi_1$. The map between the coordinates $(\rho,\phi)$ of \KutasovFR\ and $(r,\theta)$ here is
\eqn\mapdof{\rho=r\cos\theta;\qquad \phi=r\sin\theta.}

The action \dbitwo\ has the TDBI form that was studied in the context of open string tachyon condensation \refs{\SenNF\SenAN\SenTM-\KutasovER,\BergshoeffDQ\KutasovDJ\KutasovCT-\SenCZ}, 
\eqn\dbithree{\SS =-  \int d^{d+1}xV(T)\sqrt{-G}=- \int d^{d+1}x\sqrt{-g}V(T)\sqrt{1 + g^{MN}\partial_MT\partial_NT}, }
where 
\eqn\ggmmnn{G_{MN}=g_{MN}+\partial_M T\partial_N T}
is the induced metric, $G=\det\,G_{MN}$, $T=L\theta$ is the ``tachyon'' field, and 
\eqn\expvt{V(T)=\left(\cos {T\over L}\right)^{n-1}}
is the tachyon potential. Note that the tachyon field is bounded, $|T|\le T_{IR}=L\pi/2$. The potential \expvt\ vanishes at the edges of field space, a fact that will play an important role in our discussion. The mass of the tachyon, obtained by expanding \expvt\ to quadratic order and writing $V(T)=1+\half m^2 T^2+\cdots$, is
\eqn\mass {m^2 = -\frac{n-1}{L^2}.}
As a check, according to the AdS/CFT dictionary, the asymptotic behavior of a Klein-Gordon field  of mass $m$ is $r^{-\Delta}$, where $\Delta(\Delta-d) = m^2L^2$. In  \KutasovFR\ we found the large $r$ (or small $T$) behavior $T \sim r^{-d/2\pm i\sqrt{\kappa}}$, with $\kappa=n-1-d^2/4$, which agrees with \mass. 

To summarize, we see that the discussion of \KutasovFR\ can be phrased in terms of the TDBI action \dbithree. The microscopic model fixes the tachyon potential \expvt. The potential varies between one at $T=0$, which corresponds to the $Dp$ brane located at $\phi_i =0$ in the parametrization \oldmetric, and zero at $T=T_{IR}$. 

By mapping the analysis of \KutasovFR\ to the tachyon DBI language, we can deduce several properties of the Lagrangian \dbithree. The vacuum of the theory depends on whether the mass of the tachyon, \mass, is above or below the BF bound. For 
\eqn\defbf{m^2>m_{BF}^2=-\frac{d^2}{4L^2},} 
the vacuum corresponds to $T=0$ and preserves conformal symmetry. For smaller $m^2$, the vacuum is described by a non-trivial $T(r)$ which can be obtained by minimizing the energy \dbithree.  To do this, one has to introduce a UV cutoff. In \KutasovFR\ the cutoff was introduced by picking a maximal value of $\rho\le\Lambda$ and setting $\phi(\rho=\Lambda)=0$. In the TDBI coordinates, this corresponds to setting $r\le\Lambda$ and imposing the UV boundary condition $T(r=\Lambda)=0$. 

Since the equation of motion for $T(r)$ that follows from \dbithree\ is second order, we need one more boundary condition to fix the solution. In \KutasovFR\ we required that the brane be smooth at small $\rho$, which led to $\partial_\rho\phi(\rho=0)=0$. With these boundary conditions, we found that the lowest energy state of the brane has the shape depicted in figure 2. This configuration breaks conformal symmetry and dynamically generates the mass scale $\mu$. $\mu$ can be computed in terms of the UV cutoff and the distance from the transition $\kappa$; it satisfies Miransky scaling. One can take the double scaling limit $\Lambda\to\infty$, $\kappa\to 0$ with $\mu$ held fixed, which was referred to in \KutasovFR\ as the BKT limit (using the terminology of \KaplanKR). This limit focuses on the physics of the conformal phase transition while decoupling the dynamics associated with the UV cutoff.  

\ifig\loc{The shape of the brane parametrized in terms of $(\rho,\phi)$ and $(r,\theta)$.}
{\epsfxsize2.7in\epsfbox{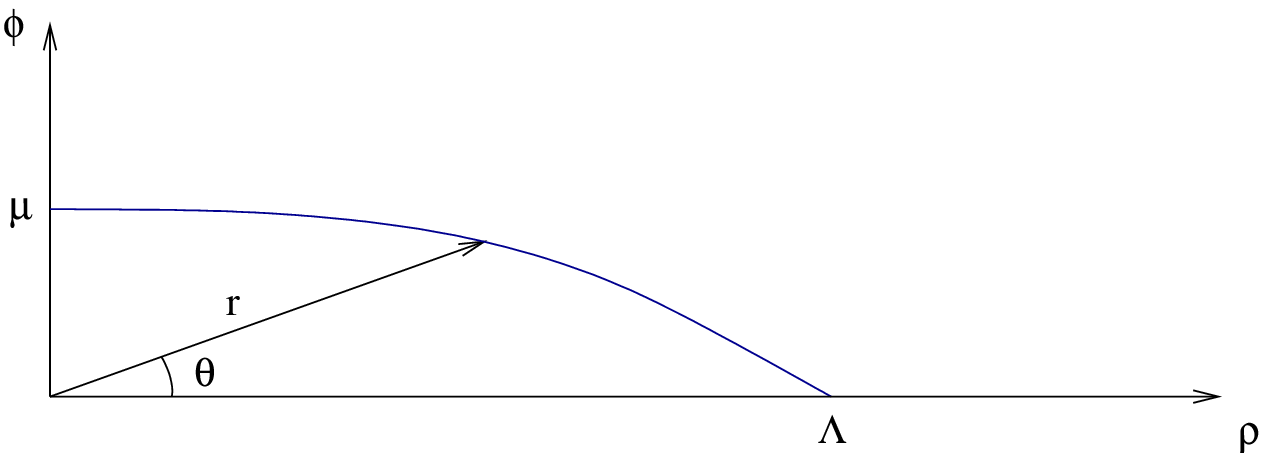}}
\ifig\loc{The shape of the brane in figure 2 parametrized in terms of $T(r)$.}
{\epsfxsize2.7in\epsfbox{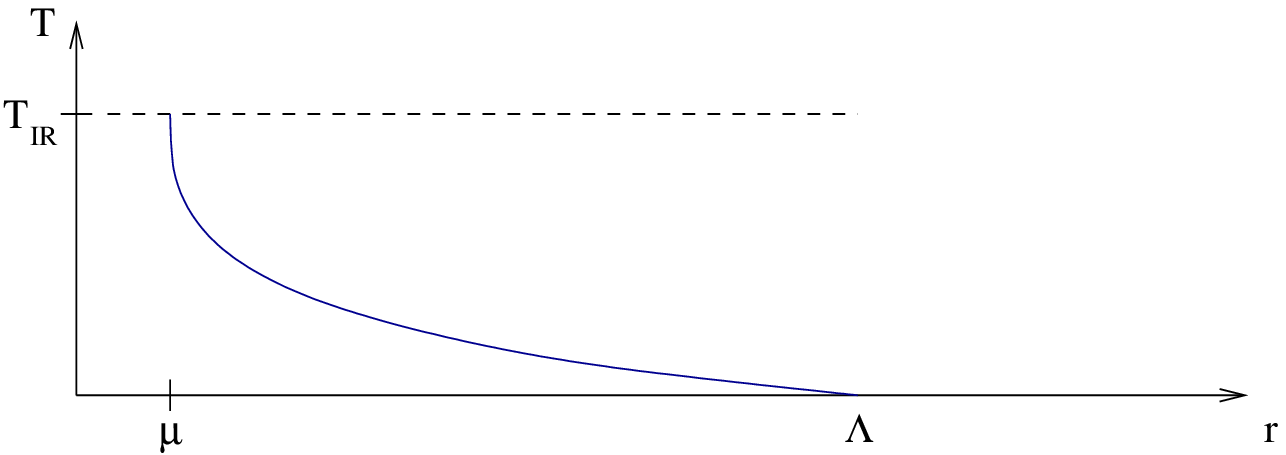}}

In terms of the TDBI variables $T$, $r$,  the shape of the brane is depicted in figure 3. Note that: 

\item{(1)} In the trivial vacuum $T(r)=0$ which preserves conformal symmetry, $r$ ranges from zero to infinity. In contrast,  in the vacuum with broken symmetry the radial coordinate is bounded from below ($r\ge \mu$). This is reminiscent of what happens in holographic models of confinement, but there the excision of a finite region in $r$ is usually put in by hand, as in the hard wall model of AdS/QCD \ErlichQH, or arises due to closed string (gravitational) dynamics \refs{\WittenZW,\KlebanovHB}. Here, the effect arises due to the dynamics of IR unstable open string modes.\foot{See also comment (4) below.} 
\item{(2)} In the $(\rho,\phi)$ parametrization used in \KutasovFR, the fact that the dynamically generated mass $\mu$ could be made arbitrarily small by tuning $\kappa$ meant that the field $\phi(\rho)$ describing the vacuum with broken symmetry  could be made arbitrarily small everywhere (see figure 2). In the $(r,T)$ parametrization the picture is somewhat different. No matter how small $\mu$ is, the field $T(r)$ always interpolates between $T=0$ at $r=\Lambda$ and $T=T_{IR}$ (recall that $T_{IR} = L\pi/2$ is the zero of $V(T)$) at $r=\mu$. The scale $\mu$ controls the size of the excised region in $r$ space, rather than the range of values taken by $T(r)$. Of course, as $\mu$ decreases, the profile $T(r)$ remains small in a larger range of $r$.
\item{(3)} The IR boundary condition $\phi'(\rho=0)=0$ translates in the $(r,T)$ parametrization to the condition that $r(T)$ is stationary at $T_{IR}$, 
\eqn\bcir{{\partial r\over\partial T}(T=T_{IR})=0,} 
or equivalently, $\partial_r T$ diverges as $T\to T_{IR}$. $r(T,x^a)$, thought of as a field living on the interval $0\le T \le T_{IR}$, satisfies Dirichlet boundary conditions at $T=0$ and Neumann ones at $T=T_{IR}$.
Thus, the scale  $\mu=r(T_{IR})$, which governs the size of the region in $r$ that is excised from the space, is dynamical. 
\item{(4)} The preceding discussion is reminiscent of what happens in the holographic models of spontaneous symmetry breaking studied in \refs{\AntonyanVW\AntonyanQY-\AntonyanPG} following \SakaiCN. Indeed, if we rotate figure 3 by ninety degrees, so that the $T$ and $r$ axes become horizontal and vertical respectively, we arrive at a similar picture to those papers. 
The symmetric configuration $T(r)=0$ can be thought of as describing a brane extending along the $r$ axis from $r=\Lambda$ to $r=0$, then going up the $T$ axis, from $T=0$ to $T=T_{IR}$. For $m^2<m_{BF}^2$ this configuration is unstable and is dynamically deformed to that of figure 3, which looks like half of the U-shape describing the symmetry breaking phase in \refs{\AntonyanVW\AntonyanQY-\AntonyanPG}. The boundary condition \bcir\ is very natural from this point of view. It ensures that one can continue the configuration of figure 3 past $T_{IR}$ and connect it to its mirror image, obtained by reflecting the configuration of figure 3 about the dashed line. This gives a smooth U-shape, very similar to those found in the above papers. Of course, there are important differences between the systems. In particular, comparing the action \dbithree\ to the one given by eq. (2.11) in \AntonyanPG, we see that the analog of $V(T)$ is constant there, while the function of $r$ in front of the square root, which here is $r^{d-1}$,  is in general different there. This accounts for the different physics of these systems. 

\noindent
An interesting question is whether the analysis of this section can be extended to more general CFTs  which  generate a mass scale due to the dynamics of operators with dimension close to $d/2$. In holography, such operators correspond to scalars with mass close to the BF bound, and if they come from the open string sector, it is natural to try to describe them by a tachyon DBI action with some $V(T)$. In the next section we will develop such a description, and study the resulting models for different $V(T)$.

\newsec{TDBI description of dynamical symmetry breaking}

Consider a $d$-dimensional CFT which contains an operator $\OO(x^a)$ with scaling dimension close to $d/2$. If this CFT has a holographic description, the operator $\OO$ is dual to a scalar field $T(x^M)$ in $AdS_{d+1}$, whose mass is close to the BF bound \defbf. Suppose that the mass depends on a tunable parameter. Then it may happen that as the parameter is varied, the mass approaches the BF bound and crosses it. In the CFT, this corresponds to the dimension of $\OO$ approaching $d/2$ and going into the complex plane. While such behavior may appear strange at first sight, it is realized in the system studied in \KutasovFR\ and is believed to be realized in (large $N$) QCD. It was also discussed in the context of the AdS/CFT correspondence in  \refs{\AdamsJB\DymarskyUH\DymarskyNC-\PomoniDE} and in the presence of background fields that break Lorentz symmetry in \refs{\JensenGA\IqbalEH\EvansHI\PalGJ\JensenVX\JensenAF-\IqbalAJ}. 

In the systems studied in \refs{\AdamsJB\DymarskyUH\DymarskyNC-\PomoniDE} the bulk field $T$ dual to $\OO$ is a closed string tachyon, while in the others it comes from the open string sector. As mentioned in the introduction, some aspects of the dynamics of open string tachyons are well described qualitatively, and in some cases quantitatively, by the tachyon DBI action. It is thus natural to ask whether the TDBI action can describe the dynamics associated with more general CPTs than the one discussed in \KutasovFR\ and the previous section. In this and the following sections, we will study such models. 

In the system discussed in the previous section, the ``tachyon'' was a geometric mode, and the ``tachyon DBI'' action was just the usual DBI action for the brane system. In the literature on open string tachyon dynamics, there have been proposals that one can more generally interpret the open string tachyon geometrically (see \eg\ \refs{\KutasovDJ\KutasovCT-\SenCZ}). Although such an interpretation fits naturally with our results, we will not assume it here.

The starting point of our discussion is the TDBI action \dbithree.  This action describes a real tachyon field $T$ on $AdS_{d+1}$, whose dynamics depends on the choice of a potential $V(T)$. We will assume that the field has a mass slightly below the BF bound, \ie\ the potential $V(T)$ has a local maximum at the origin:
\eqn\vtsmall{V(T)=1+\half m^2T^2+\cdots }
where $m^2<m_{BF}^2=-d^2/4$.\foot{Here and below we take the $AdS$ radius to be $L=1$.}  We will further assume that $V(T)$ is a monotonically decreasing function of $T$ that goes to zero at a particular value of $T$, $T=T_{IR}$. As we will see shortly, there is a qualitative difference between the cases of finite and infinite $T_{IR}$ (see figure 4). In the system of \KutasovFR\ $T_{IR}$ was finite, while for the potentials that arise in open string tachyon condensation it is typically infinite \refs{\MinahanTF,\KutasovAQ}. We will consider both cases below.   

\ifig\loc{The tachyon potential, for finite (a) and infinite (b) $T_{IR}$.}
{\epsfxsize2.5in\epsfbox{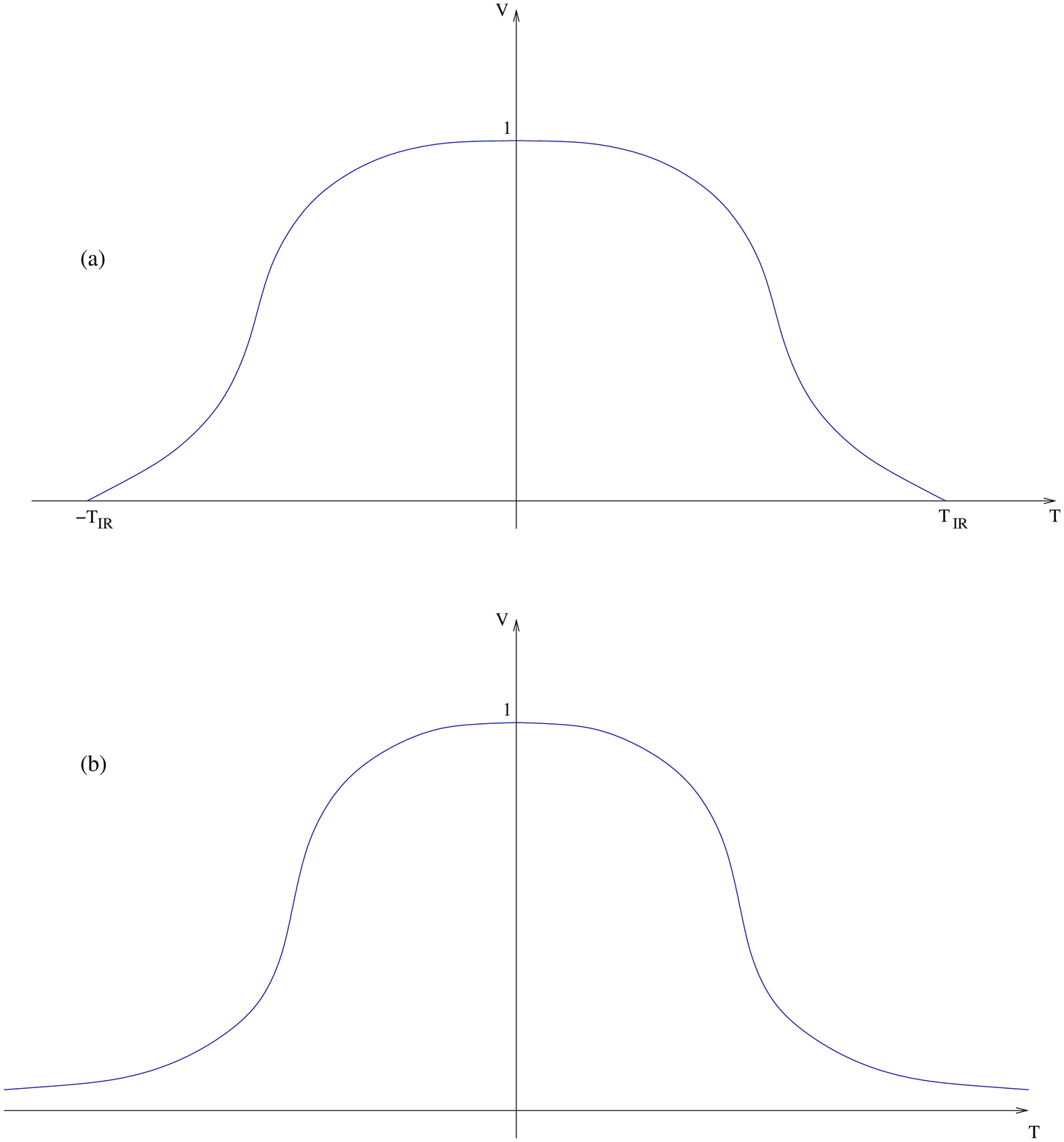}}

The fact that the potential vanishes somewhere in field space is important. In the study of open string tachyon condensation this value of the tachyon corresponds to the state in which the unstable D-brane (or brane-antibrane pair) disappears \SenNF.  In the defect fermion system of \KutasovFR\ it is the state where the probe D-brane is pushed to infinite $\phi$, so that the fermions have an infinite current mass and decouple. Such a decoupling limit should exist for other systems  described by \dbithree\ as well. 

One can think of the TDBI action as an analog of the Landau-Ginzburg action for an order parameter in the vicinity of a first or second order phase transition. Here, the order parameter $T$ is a bulk field, which describes an infinite number of light states in the boundary theory. The potential $V(T)$ labels different ``universality classes'' of such transitions.  

Before turning to a more detailed investigation of the action \dbithree, we comment on some generalizations. If the boundary CFT we are studying has a $U(1)$  global symmetry (under which the operator $\OO$ is not charged), generated by a conserved current $J^a$, we need to include in the bulk description a gauge field $A^M$. This can be achieved by replacing $G_{MN} \rightarrow G_{MN} + F_{MN}$ in \dbithree. The modified action can be used to calculate correlation functions of the current $J^a$ and the spectrum of vector mesons, and to turn on a chemical potential for the corresponding conserved charge.  

In QCD, the tachyon field is complex, like the dual fermion bilinear $\psi_L^\dagger\psi_R$.  This can be accomodated by using the TDBI action for the $D-\bar D$ tachyon  \refs{\SenNF,\SenTM}, 
\eqn\complextdbi{\SS =-  \int d^{d+1}xV(T^\dagger T)\left(\sqrt{-G^{(L)}}+\sqrt{-G^{(R)}}\right)}
where 
\eqn\openmetric{
G_{MN}^{(L)} = g_{MN} + \frac 12 D_{(M}T^\dagger D_{N)}T+F_{MN}^{(L)}}
and similarly for $L\leftrightarrow R$. In QCD with $N_f$ flavors, the tachyon transforms in the $(N_f,\bar{N_f})$ representation of the $SU(N_f)_L\times SU(N_f)_R$ global symmetry group. DBI-type actions for matrix fields are known to be inherently ambiguous. For the purpose of our discussion, it is enough to study the full non-perturbative action \complextdbi\ for  the trace of the $N_f\times N_f$ matrix $T$, which develops a non-trivial vacuum expectation value. The rest of the components of $T$, as well as the $SU(N_f)_L\times SU(N_f)_R$ gauge fields $A_M^{(L,R)}$ which are dual to the global symmetry currents, can be treated perturbatively. In particular, to study the spectrum of mesons we only need to work to quadratic order in these fields, in which \complextdbi\ is unambiguous. 

The vacuum of the model \dbithree\ is described by specifying a field configuration $T=T_0(r)$ which minimizes the energy function obtained from \dbithree,
\eqn\expaction{
\EE = \int dr r^{d-1}V(T)\sqrt{1+r^2T'(r)^2}.}
The equation of motion following from \expaction\ can be written as
\eqn\eomttt{
\frac{r^{d-1}\partial_T\ln V}{\sqrt{1+r^2T'^2}} = \frac{\partial}{\partial r}\left( \frac{r^{d+1}T'}{\sqrt{1+r^2T'^2}}\right),
}
or equivalently, 
\eqn\eomtwo{-(1+r^2T'^2)\partial_T\ln V + (1+d)rT' + dr^3T'^3 + r^2T''=0.}
We believe, but have not proven in general, that for any potential of the qualitative form in figure 4, the vacuum is trivial (\ie\ $T_0(r) = 0$ for all $r$) when the mass \vtsmall\ is above the BF bound. We will mention some evidence supporting this claim below. We will also see that for $m^2$ below the BF bound the vacuum corresponds to a non-trivial solution, $T_0(r)$, which takes the qualitative form depicted in figure 3. The boundary conditions satisfied by the solution are the same as those of section 2: $T(r=\Lambda)=0$ in the UV, and \bcir\ in the IR. For models where $T_{IR}\to\infty$, there is no excised region and $T_0(r)$ takes the form shown in figure 5. It can be thought of as a limit of the solution in figure 3, obtained by taking $T_{IR}\to\infty$ and $\mu\to 0$. 

\ifig\loc{The vacuum configuration $T_0(r)$ for infinite $T_{IR}$.}
{\epsfxsize2.7in\epsfbox{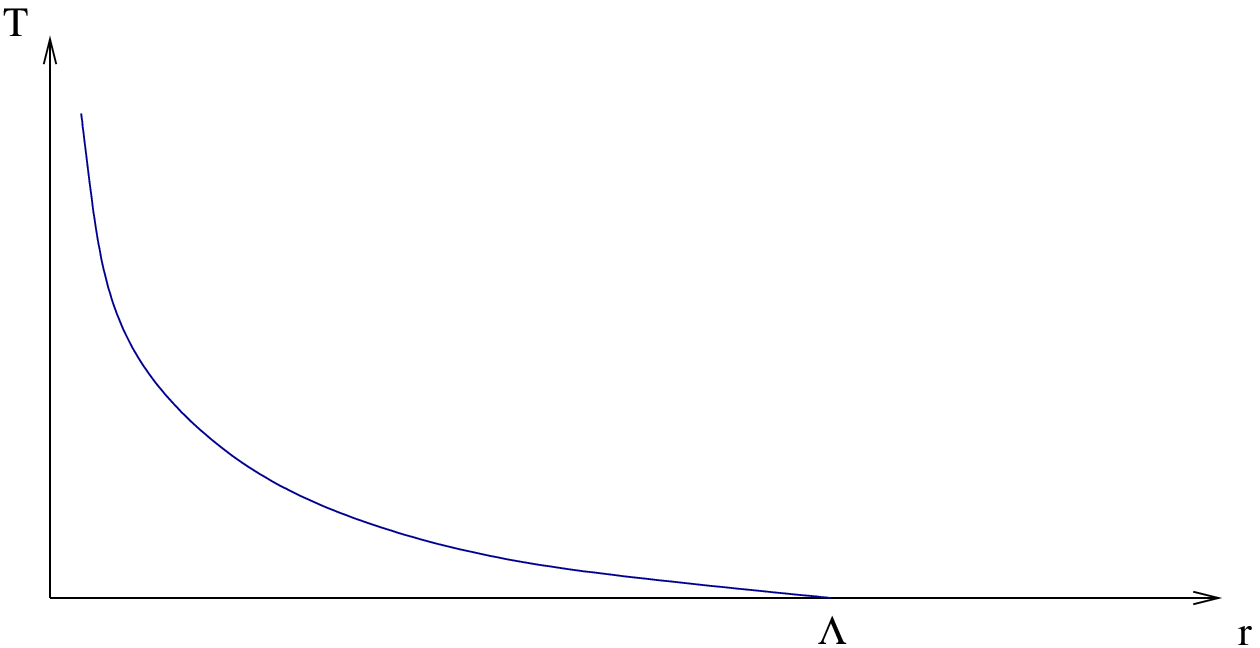}}

As is clear from figures 3 and 5, at large $r$ the solution of \eomtwo\ is small, and we can replace the full equation of motion by the linearized one, 
\eqn\lineom{r^2T''+(1+d)rT'-m^2T=0.}
The general solution of this equation is 
\eqn\asymptsol{T_0(r)=A\left(\mu\over r\right)^{d\over2}\sin\left(\sqrt\kappa\ln{r\over\mu}+\phi\right)}
where 
\eqn\defkappa{\kappa=m_{BF}^2-m^2}
is assumed to be positive, $\mu$ is a scale, and $A$, $\phi$ are dimensionless constants whose  definition depends on $\mu$. As $r$ decreases, $T_0(r)$ increases, until eventually the linear approximation \lineom\ breaks down and one has to go back to the full equation of motion. $\mu$ can be thought of as the scale at which this happens. The advantage of this definition is that it applies both to models with finite $T_{IR}$, where $\mu$ here is of the order of the scale $\mu$ defined in figure 3, and with infinite $T_{IR}$, where there is no excised region. Of course, this definition is not precise, but it can be made precise \eg\ by setting $\mu$ to be the mass of the lowest lying meson. Other definitions differ from this one by numerical factors. The ambiguity in the definition of the dynamically generated scale is familiar in QFT, and it can be dealt with using the renormalization group. 

In the BKT limit $\Lambda\to\infty$, $\kappa\to 0$ with $\mu$ held fixed, the large $r$ behavior of $T_0(r)$ becomes \KutasovFR\ 
\eqn\bktlarger{T_0(r)=\left(\mu\over r\right)^{d\over2}\left(C_1\ln{r\over\mu}+C_2\right)}
where $C_1$, $C_2$ are again dimensionless constants, related to $A$, $\phi$ as follows:
\eqn\aphicc{C_1=A\sqrt\kappa;\qquad C_2=A\phi.}

The presence of a non-trivial tachyon condensate $T_0(r)$ leads to a deformation of the $AdS_{d+1}$ metric $g_{MN}$ to 
\eqn\opengmn{G_{MN}=g_{MN}+\partial_MT_0\partial_N T_0}
and to a non-trivial dilaton $\Phi(r)$. Excitations described by the TDBI action live in this modified, or open string, background. As $r\to\infty$, the open string metric $G$ approaches the closed string (AdS) metric $g$ and the dilaton approaches a constant, but for small $r$, the open and closed string backgrounds are in general quite different. In the rest of this section, we will derive the equations describing fluctuations of the TDBI action around the vacuum $T_0(r)$.

We start with fluctuations of the field $T(x^M)$, which give rise to $\sigma$-mesons. We define the fluctuation $y$ via $T(x^M)=T_0(r)+y(x^M)$, plug this into \dbithree, and expand to quadratic order in $y$. The resulting action for the fluctuations is
\eqn\mesonaction{
\SS_2 = -\half\int d^dx dr \sqrt{-G}e^{-\Phi}\left(G^{MN}\partial_My\partial_Ny + m^2(r)y^2\right)}
where 
\eqn\mcoeff{\eqalign{
e^{-\Phi} =& \frac{V(T_0)}{(1+r^2T_0'^2)}, \cr
m^2(r) =& \frac{\sqrt{1+r^2T_0'^2}}{V}\left[\frac{\partial^2V}{\partial T^2}\frac{1}{\sqrt{1+r^2T_0'^2}} - (d-1)\frac{\partial V}{\partial T}\frac{rT_0'}{\sqrt{1+r^2T_0'^2}}- \frac{\partial V}{\partial T}\frac{\partial}{\partial r}\left(\frac{r^2T_0'}{\sqrt{1+r^2T_0'^2}} \right) \right],
}}
and $G_{MN}$ is the open string metric \opengmn.
It is easy to check using \vtsmall, \bktlarger, that at large $r$ the dilaton $\Phi$ approaches a constant, and the mass function $m^2(r)$  goes to the tachyon mass $m^2$. 

To study the spectrum of mesons, it is convenient to change coordinates from $r$ to $z(r)$, defined by the requirement that the open string metric $G_{MN}$ takes the form 
\eqn\newcoords{G_{MN} dx^Mdx^N=h_{\alpha\beta}dx^\alpha dx^\beta= r^2(z)dx^\alpha dx^\beta\eta_{\alpha\beta}.}
Here $(x^\alpha)=(z, x^a)$. Comparing \opengmn\ and \newcoords\ we see that the coordinates $z$ and $r$ are related as follows:
\eqn\cov{
\frac{dz}{dr} = -\frac{\sqrt{1+r^2T_0'^2}}{r^2}. }
For large $r$, \cov\ reduces to $z=1/r$, and \newcoords\ becomes a familiar parametrization of the metric on $AdS_{d+1}$. For small $r$, there are significant deviations from $AdS$, which we will discuss below. 

The quadratic action for fluctuations \mesonaction\ is given in these coordinates by 
\eqn\naction{\eqalign{
\SS_2 =& -\half\int d^{d+1}x \sqrt{-h}e^{-\Phi}\left(h^{\alpha\beta}\partial_\alpha y\partial_\beta y + m^2(r(z))y^2\right)\cr
=&-\half\int d^{d+1}x r^{d-1}e^{-\Phi}\left(\partial_\alpha y\partial^\alpha y + m^2(r)r^2y^2\right)}}
where on the second line $r$ is viewed as a function of $z$, and indices are raised and lowered with the flat metric $\eta_{\alpha\beta}$. We next define the wavefunction $\psi = ye^{-B/2}$ where
\eqn\bz{
B(z) = \Phi(z) - (d-1)\ln r(z),}
and write it as $\psi(x^\alpha)=\psi(z)\exp(ik_ax^a)$. The equation of motion for $\psi(z)$ that follows from \naction\ takes the 
Schr\"odinger form
\eqn\schrodinger{
-\psi'' + V_{\rm{eff}}(z)\psi = \MM^2\psi}
where the prime denotes differentiation w.r.t. $z$, $\MM^2=-k_ak^a$ is the meson mass, and the effective potential is 
\eqn\effpot{
V_{\rm eff}(z) = \frac 14(B')^2 - \frac12 B'' + r(z)^2m^2(r(z)).}

The large $z$ (IR) behavior of the potential $V_{\rm eff}(z)$ depends on the details of the tachyon potential $V(T)$. For small $z$, however, the potential is universal: $T_0(r)$ goes to zero as $r\to\infty$, the dilaton $\Phi(r)$ and mass function $m(r)$ approach constants, and $r=1/z$. Thus, in this regime one finds $B(z)=(d-1)\ln z$ and  
\eqn\veffir{
V_{\rm eff}(z) = \left(\frac{d^2-1}{4} + m^2\right)\frac{1}{z^2} = -\left({1\over4}+\kappa\right){1\over z^2}.
}
We are interested in $\kappa>0$ (the massive phase), in the presence of a UV cutoff, which provides a lower bound $z\ge z_{UV}=1/\Lambda$  on $z$. The boundary condition there is $\psi(z_{UV})=0$. If the potential $V_{\rm eff}$ was given by \veffir\ for all $z\ge z_{UV}$, the spectrum of \schrodinger\ would contain normalizable states with $\MM^2<0$, \ie\ tachyons. The mass squared of the lowest lying of these tachyons is of order $-\mu^2$, where $\mu\sim\Lambda\exp(-\pi/\sqrt\kappa)$. It is easy to understand this from the point of view of the TDBI system. The potential \veffir\ describes small fluctuations about the conformally invariant state $T(r)=0$. This state is unstable to condensation of $T(r)$ towards the configuration $T_0(r)$ described in figures 3 and 5. The tachyons of \veffir\ are the eigenmodes of this condensation process. 

When studying perturbations around the stable configuration $T_0(r)$, the potential takes the form \veffir\ for $z\ll1/\mu$, but is deformed at $z\sim 1/\mu$ and larger. Since the wavefunctions corresponding to the tachyons  of \schrodinger, \veffir\ have finite support at $z\sim1/\mu$, the spectrum of $V_{\rm eff}$ in the state corresponding to $T_0(r)$ is deformed relative to that of the configuration $T(r)=0$ (even in the BKT limit). In particular, it cannot have tachyonic modes by construction, since it is the lowest energy state with the prescribed boundary conditions. 

Notice that for $\kappa<0$, 
 the spectrum of \schrodinger, \veffir\  does not contain any bound states which would correspond to tachyons. This supports our earlier assertion that in the conformal phase, the state $T(r)=0$ is the ground state of the system -- it shows that $T(r)=0$ is at least locally stable. The fact that $\kappa=0$ lies on the boundary between two different behaviors of the bound state problem corresponding to the potential \veffir\ is familiar from the QM analysis of this potential (see \eg\  \LLL).

To study the spectrum of vector mesons, we have to add a gauge field $A_M$ to the TDBI action. The action then becomes%
\eqn\gaction{
\SS =- \int d^{d+1}xV(T)\sqrt{-\det (G_{MN} +F_{MN})} \simeq   \SS_0-\frac14\int d^{d+1}x\sqrt{-G}V(T)G^{MM'}G^{NN'}F_{MN}F_{M'N'} .
}
Thus, the gauge field propagates in a spacetime with metric $G_{MN}$ \opengmn, and dilaton 
\eqn\gaugedil{e^{-\Phi}=V(T_0).}
It is interesting that while the metric that the vectors and scalars experience is the same, the dilaton is not -- it is given by \mcoeff\ for scalars and \gaugedil\ for vectors. Another difference is that the analog of $m(r)$ in \mesonaction\ vanishes for the gauge field. 

It is convenient to change coordinates as in \newcoords, \cov, choose the gauge $A_z=0$, and write the remaining components of the gauge field as $A_a=\xi_a(x^b)\exp(B/2)\psi(z)$, where $\xi_a$ is the (transverse) polarization of a gauge field of momentum $k^a$. The wavefunction $\psi(z)$ then satisfies a Schr\"odinger-type equation of the form \schrodinger. $V_{\rm eff}$ is given by \effpot\ with $m(r)=0$ and 
\eqn\gcoeff{B(z) =- \ln V(T_0) -(d-3) \ln r(z).}
The small $z$ (UV) behavior of the potential $V_{\rm eff}$ is again insensitive to the details of the tachyon potential; it is $V_{\rm{eff}} = (d-1)(d-3)/(4z^2)$. Thus we see that in the UV, the potentials for the $\sigma$ and $\rho$ mesons both go like $V_{\rm{eff}}=-\alpha/z^2$, but while $\alpha_\sigma$ takes the critical value $1/4$, $\alpha_\rho$ is strictly below this value. This may be related to the fact that while the $\sigma$-meson spectrum was found in \KutasovFR\ to contain a light mode which can be thought of as a pseudo-Nambu-Goldstone meson of broken conformal symmetry, the $\rho$-mesons were all heavier. 

One also needs to check whether fluctuations of the gauge field $A_M$ give rise to scalars. In the gauge $A_z=0$, the scalar modes may be parametrized as $A_a(z,x^a)=\phi(z)\partial_a\pi(x^b)$. It is easy to check that the equations of motion of \gaction\ do not allow such solutions with non-zero mass. The case of zero mass was analyzed in \KarchEG, where it was shown that a massless meson appears if the integral 
\eqn\nnnmmm{N_m=\int dz e^B=\int {dz\over V(T) r^{d-3}}}
is finite. At small $z$, one can set $V(T)\sim 1$, $r\sim 1/z$, so the contribution from this region is finite (when $d>2$). Thus, the fate of the massless meson lies in the large $z$ region. We will return to it in later sections and see that for sensible potentials, the normalization integral \nnnmmm\ is divergent, so there is no massless scalar in this sector.

So far, we have discussed the physics associated with the TDBI action of a real tachyon \dbithree. In QCD the tachyon is complex, and it is natural to study the action \complextdbi\ describing a complex tachyon and gauge fields $A^{(L)}$, $A^{(R)}$ that couple to the chiral currents. In this case it is natural to write the complex tachyon field as $T=\tau\exp(i\theta)$, and define vector and axial gauge fields $V=A^{(L)}+A^{(R)}, A=A^{(L)}-A^{(R)}.$
The vacuum solution can be taken to be $\theta=0$, $V=A=0$, and $\tau=\tau(r)$. Then the complex action \complextdbi\ reduces to the TDBI action for a real tachyon field, and $\tau(r)$ coincides with $T_0(r)$ defined earlier in this section.

In this system, there are four types of small fluctuations which give rise to mesons. Scalar and vector mesons correspond to fluctuations of the fields $\tau$ and $V$, and are governed by the same equations as in the real TDBI case. Axial vector mesons arise from transverse fluctuations of the gauge field $A$, and are described by the quadratic action
\eqn\gactiontwo{
\SS_2 \simeq  -\int d^{d+1}x\sqrt{-G}V(\tau)\left(\frac 18G^{MM'}G^{NN'}F^{(A)}_{MN}F^{(A)}_{M'N'} + G^{MN}\tau^2 A_MA_N\right).
}
The first term is the same as in \gaction\ (up to a coefficient from the way we defined the vector and axial fields here). The second is due to the fact that $T$ is charged under the axial gauge field. Following the same steps as above (which are discussed in detail in Appendix A), we find that axial vector mesons are obtained by solving the Schr\"odinger problem 
\schrodinger\ with the effective potential 
\eqn\schraxial{V_{\rm{eff}}^{(A)} =  \frac 14(B')^2 - \frac12 B'' + (2r\tau(r))^2.} 
Here $B(z)$ is given by \gcoeff, and $r(z)$ is the solution of \cov, as before. The difference between the vector and axial case lies solely in the last term in \schraxial. It is negligible in the UV, but in general important in the IR. Pseudoscalars arise from longitudinal fluctuations of the axial gauge field and fluctuations of $\theta$, the phase of $T$. We  discuss them in Appendix A. 

In this section, we presented the general structure of the vacuum and excitations of the tachyon DBI system in the vicinity of the CPT. We next analyze some aspects of the dynamics for different classes of potentials. We start in the next section with a discussion of potentials $V(T)$ which vanish at a finite value of $T$, and  then move on to those that asymptote to zero as $T\to\infty$.

\newsec{Hard wall models}

In this section, we consider models in which $T_{IR}$ is finite and the tachyon potential $V(T)$ takes the form in figure 4(a). We will refer to such models as hard wall models, for reasons that will become clear shortly. We assume that as $T\to T_{IR}$, the potential behaves like 
\eqn\irpoten{V(T)\simeq (T_{IR}-T)^\alpha}
where $\alpha$ is a positive real number. This includes the defect fermion system of section 2, but is of course not the most general behavior. It is easy to generalize our discussion to other cases, but we will not do so here.

Plugging \irpoten\ in the equation of motion \eomtwo, one can check that any solution that reaches $T=T_{IR}$ behaves in its vicinity like 
\eqn\behavtzero{T_{IR}-T_0(r)\simeq c\sqrt{r-\mu}; \qquad  c = \sqrt{\frac{2(\alpha+1)}{d\mu}},}
where  $\mu$ is a dynamical scale that can be calculated by matching the behavior near $T_{IR}$ to the behavior \asymptsol\ near the UV cutoff.  Note that the solution \behavtzero\ satisfies the boundary condition \bcir. Indeed, the function $r(T)$ has a quadratic minimum at $T_{IR}$, $r(T)-\mu=(T-T_{IR})^2/c^2$. 
As $T$ varies between zero and $T_{IR}$, $r$ varies between $\Lambda$ and $\mu$. Thus, fluctuations described by the TDBI action see a space cut off in the IR, as in hard-wall AdS/QCD \refs{\ErlichQH,\DaRoldZS} and the models of \refs{\AntonyanVW\AntonyanQY-\AntonyanPG}. 

To study the spectrum of mesons, it is convenient to change coordinates from $r$ to $z$ using \cov. Since $z$ is a decreasing function of $r$, an important question is whether the region $r\ge\mu$ is mapped to a compact region in $z$. It is easy to see that the answer is affirmative. Indeed, when $r\simeq\mu$, one can replace the factors of $r$ in \cov\ by $\mu$ and use the fact that $T_0'(r)\to\infty$ as $r\to\mu$ to rewrite  \cov\ as 
\eqn\zuvv{{d\over dr}\left(\mu z-T_0\right)=0.}
Since $T_0(r)$ approaches a finite value $T_{IR}$ as $r\to\mu$, $z(r)$ must approach a finite value $z_{IR}$ as well. To calculate this value, we need to solve the full e.o.m. \eomtwo\ and integrate \cov, which is typically only possible numerically. 

The discussion of the last paragraph is valid for any model with finite $T_{IR}$, $\mu$. Specifying to potentials of the form \irpoten\ and using \behavtzero, we find that 
\eqn\formzr{z_{IR}-z(r)={c\over\mu}\sqrt{r-\mu}.}
In section 3 we saw that scalar perturbations are described by the Klein-Gordon Lagrangian \mesonaction\ with a non-trivial metric \opengmn, dilaton and mass function \mcoeff. For potentials $V(T)$ that behave like \irpoten\ near $T_{IR}$, we can  calculate the form of the dilaton $\Phi(z)$ and mass function $m(z)$ near $z_{IR}$. They turn out to be
\eqn\formdilmass{\eqalign{
\Phi(z)=&-(\alpha+2)\ln(z_{IR}-z),\cr
m^2(z)=&-{\alpha\over\mu^2(z_{IR}-z)^2.}
}}
Plugging eqs. \formzr, \formdilmass\ into  \effpot, we find that the effective potential for $\sigma$-mesons is
\eqn\effpotsig{V_{\rm eff}(z)\simeq {\alpha(\alpha-2)\over4}{1\over (z_{IR}-z)^2}}
near $z=z_{IR}$. The spectrum of $\sigma$-mesons is obtained by solving the Schr\"odinger equation \schrodinger, with a potential that behaves like \veffir\ for small $z$ and \effpotsig\ for $z\sim z_{IR}$. The boundary conditions for the wavefunction $\psi$ are Dirichlet at $z=0$ and Neumann at $z=z_{IR}$. 

It is interesting that the potential $V_{\rm eff}$ diverges like $A/(z-z_b)^2$ near both boundaries $z_b=0, z_{IR}$. As is well known (see \eg\ \LLL), for such potentials the coefficient  $A$ is important. Near $z=0$, this coefficient takes the critical value $A=-1/4$ separating different regimes; near $z_{IR}$ it is at the critical value when $\alpha =1$ and otherwise strictly above it. In fact, for $\alpha>2$ the potential goes to infinity as $z\to z_{IR}$. 

To solve for the full spectrum of $\sigma$-mesons requires knowledge of the full potential $V_{\rm eff}$ which interpolates between the asymptotic behaviors \veffir\ and \effpotsig.  However, for highly excited states one can neglect the potential, and read off the spectrum from the locations of the walls. This gives 
\eqn\mmmnnn{m_n\simeq {\pi n\over z_{IR}}.}
For the defect example discussed in \KutasovFR\ and section 2, one can solve  \cov\ numerically, and show that $z_{IR}\sim 1.6/\mu$. Plugging into \mmmnnn\ gives the asymptotic spectrum 
\eqn\asymspec{m_n\simeq 1.96\mu n,}
in good agreement with eq. (4.11) in \KutasovFR, $m_n\simeq 1.97 \mu n$. 

It is easy to repeat this discussion for vector mesons. The dilaton is given in this case by \gaugedil, which behaves near $z_{IR}$ as
\eqn\effgdil{\Phi(z)=-\alpha\ln(z_{IR}-z).}
Plugging into \effpot\ (with $m(r)=0$), we find that the effective potential for vector mesons behaves near $z_{IR}$ like \effpotsig\ as well. Hence, we expect the spectrum of vector mesons to have the same asymptotics as the scalar spectrum. Eq. (4.22) in \KutasovFR\ gives $m_n\simeq 1.96\mu n$, in excellent agreement with \asymspec.\foot{The agreement for vectors is slightly better than that for scalars since in \KutasovFR\ we deduced the asymptotic spectrum of vector mesons from more states than for scalars.}

In section 3, we mentioned that longitudinal fluctuations of the vector field can in principle support a massless scalar meson if the integral $N_m$ \nnnmmm\ is finite. The only possible divergence of this integral comes from the region $z\to z_{IR}$. Plugging \irpoten, \behavtzero, and \formzr\ into \nnnmmm, we find 
\eqn\intzr{N_m\sim \int^{z_{IR}} {dz\over (z_{IR}-z)^\alpha}.}
It is finite for $\alpha<1$, and divergent otherwise. This is a reasonable result, since for $\alpha<1$ $V'(T)$ diverges as $T\to T_{IR}$. If we demand that the gradient of  $V$ remains finite, the massless scalar excitation is non-normalizable.

To summarize, we find that any system described by the TDBI action \dbithree, with a tachyon potential $V(T)$ that behaves as \vtsmall\ with $m$ slightly below the BF bound for small $T$, and like \irpoten\ near the point $T_{IR}$ where the potential vanishes, dynamically breaks conformal symmetry. The vacuum is described by a configuration $T=T_0(r)$, which behaves like \asymptsol,  \bktlarger\ for large $r$ and like \behavtzero\ near $T_{IR}$. The spectrum of scalar and vector mesons is discrete and has the asymptotic Kaluza-Klein form \mmmnnn. 
Although these results were obtained for potentials with the asymptotic behavior \irpoten, we believe that the qualitative picture holds much more generally for models with finite $T_{IR}$. We will leave further study of such models to future work. 

\newsec{Soft wall models}

The models described in the previous section have the property that the asymptotic high mass spectrum has the KK form \mmmnnn. In some applications, this is an undesirable feature. In particular, in large $N$ QCD it is believed that the masses of highly excited mesons behave asymptotically like
\eqn\thooftas{m_n^2\sim n,}
sometimes referred to as a linear confining spectrum. Hence, it is of interest to try to accommodate such behavior in a holographic framework.\foot{Of course, near the CPT of QCD there is no particular reason to expect the highly excited meson spectrum to have the form \thooftas, but it may play a role in this and other applications.}  In AdS/QCD, this was achieved by replacing the hard IR cutoff on the radial coordinate with a soft cutoff corresponding to a non-trivial radial dilaton profile, chosen to get linear confinement  \KarchPV. More elaborate constructions appear in  \refs{\CaseroAE,\BatellZM}.

In this section, we will see that linear confinement and other non-trivial behaviors can be achieved  in our framework  by considering potentials $V(T)$ with $T_{IR}\to\infty$. Such potentials have the qualitative form of figure 4(b), and  appear naturally in the study of open string tachyon condensation in string theory \refs{\SenNF\SenAN\SenTM-\KutasovER,\KutasovDJ\KutasovCT-\SenCZ}. We will refer to them as soft wall potentials.

As in the previous section, we will not analyze the most general possibility. As a first step, we consider the class of potentials that behave at large $T$ like 
\eqn\inftir{V(T)\simeq e^{-\half\beta T^2}}
where $\beta$ is a positive constant. Potentials of this form (with a fixed value of $\beta$) naturally arise in studies of open string tachyon condensation \refs{\MinahanTF,\KutasovAQ}. There is no particular  reason to expect them to play a role in AdS/QCD, but we will see that they give rise to interesting, $\beta$-dependent, physics. 

Proceeding as in section 4, we substitute the potential \inftir\ into the equation of motion \eomtwo\ to find the form of the vacuum solution $T_0(r)$. The small $r$ behavior of this solution is 
\eqn\smallrtzero{T_0(r)\simeq \left(\mu\over r\right)^{\beta\over d}.}
It has the qualitative form depicted in figure 5, in agreement with the discussion of section 3. In particular, there is no excised region in $r$, like in soft wall AdS/QCD \KarchPV. 

As before, we change variables from $r$ to  $z$ using the map \cov. At small $r$, this map reduces to
\eqn\zrsoft{z(r)\sim {1\over r} \left(\mu\over r\right)^{\beta\over d}. } 
Note that as $r\to 0$, $z\to\infty$, so the $z$ coordinate runs from zero to infinity as well. 

We can again compute the effective dilaton and mass function for scalar mesons \mcoeff, which (for large $z$) take the form 
\eqn\formphm{\Phi(z), m^2(z)\sim  T^2\sim (\mu z)^{2\beta\over \beta+d}.}
Plugging into \effpot, we find the large $z$ behavior of the effective potential, 
\eqn\largezveff{V_{\rm eff}\sim \mu^2(\mu z)^{2{\beta-d\over\beta+d}}.}
When $\beta<d$, the effective potential \largezveff\ goes to zero as $z\to\infty$. Hence the spectrum contains a continuum of ``scattering'' states living at large $z$. When $\beta>d$, the potential grows without bound as $z\to\infty$, and one expects a discrete spectrum. The asymptotic form of the spectrum is determined by the behavior of the potential at large $z$. One can read off the asymptotic spectrum from a similar analysis performed in \BatellME:
\eqn\asmn{m_n\sim  \mu n^{\half\left(1-{d\over\beta}\right)}.}

The above discussion was for scalar mesons, but it is easy to check that the behavior \largezveff, \asmn\ is similar for vector mesons.  The two cases differ only in the proportionality constant in \asmn.  This difference comes from the last term in the effective potential \effpot\ for scalars, which contributes to the coefficient and is absent for vectors.

In section 3 we mentioned the possibility that longitudinal vector excitations give rise to a massless scalar meson, depending on the finiteness of the normalization integral \nnnmmm. It is easy to see that for potentials with the behavior \inftir\ this integral is always divergent, so such a scalar is absent.

For axial vectors, the effective potential is given by \schraxial, which differs from the effective potential for vectors by the last term $\sim(r\tau)^2$. Using \zrsoft, \formphm\ it is easy to check that this term contributes to the coefficient of $z^{2(\beta-d)/(\beta+d)}$ in \largezveff. Thus, the asymptotic spectra of vector and axial vector mesons in models with potentials of the form \inftir\ are different.\foot{The same is true for scalars and pseudoscalars.} This is at first sight puzzling, since one might think that large energies correspond to large $r$, or small $z$, where the background $\tau(r)$ goes to zero and chiral symmetry of the Lagrangian \complextdbi\ appears to be restored. We will return to this issue and explain why this intuition fails in section 7.

We see that soft wall tachyon DBI models naturally give spectra different from those of hard wall models. Note that as $\beta\to\infty$,  \asmn\ approaches a linearly confining spectrum. Thus, it is natural to ask whether there exists a potential that reproduces the behavior \thooftas\ exactly. Presumably, as $T\to\infty$ such a potential would have to go to zero faster than \inftir\ with any finite $\beta$. 

With this in mind, we next consider a potential $V(T)$ with the large $T$ asymptotic behavior
\eqn\ttlltt{V(T)\simeq e^{-\gamma T^2\ln T}.}
Here $\gamma$ is a positive constant whose value will play a role below. As is by now familiar, we start by solving eq. 
\eomtwo\ for the large $T$ form of the vacuum configuration $T_0(r)$, which turns out to be
\eqn\tsol{T_0(r) \simeq \exp\left(\mu\over r\right)^{2\gamma\over d}. }
We make the change of variable \cov\ and find that in the IR
\eqn\formzr{z(r) \simeq {1\over r}\exp\left(\mu\over r\right)^{2\gamma\over d}=\frac{T_0(r)}{r}. }
The inverse map is $r\simeq \mu(\ln \mu z)^{-d/2\gamma}$. The dilaton behaves at large $z$ like $\Phi\simeq -\ln V(T_0)=\gamma T_0^2\ln T_0$. Using \tsol,  eq. \formzr\ gives
\eqn\fffphi{\Phi(z)\simeq\gamma \mu^2z^2 (\ln z)^{1-{d\over\gamma}}.}
When $\gamma=d$, one finds $\Phi\sim \mu^2 z^2$;  the potential $V_{\rm eff}$ \effpot\ for scalar mesons and its analog for vector mesons goes in this case like $\mu^4z^2$, which leads to the linearly confining spectrum $m_n^2\sim \mu^2 n$. For other values of $\gamma$ the quadratic potential receives logarithmic corrections, which deform the linear spectrum \thooftas.

The coefficient of $\mu^2 n$ in $m_n^2$ is different for scalar and vector mesons. This is because the mass term in the effective potential (the last term in  \effpot) contributes to the  coefficient of $z^2$ in $V_{\rm eff}$ for scalar mesons, and is absent for vector mesons. On the other hand, the difference between the highly excited spectra of vector and axial vector mesons lies the last term of the axial vector potential \schraxial, which goes like $(r\tau)^2\sim (r^2z)^2\sim z^2/(\ln z)^2$. It is (marginally) subleading at large $z$, so the spectra of vectors and axial vectors  (slowly) approach each other at high excitation levels.

To summarize, in this section we saw that potentials of the general form of figure 4(b) give qualitatively different physics than those of figure 4(a). Rather than leading to an excision of a finite region of small $r$, they provide a soft wall that smoothly suppresses dynamics at small $r$. By varying the potential, one can arrange the spectrum to have different asymptotic forms such as \thooftas\ and \asmn. All the features described above arise dynamically once a potential $V(T)$ is specified. This is in contrast to the usual approach where the profile of the dilaton and other closed string fields are put in by hand. As we will discuss in section 7, in QCD the difference between the two approaches is that one is obtained by expanding around the boundary between confining and conformal behavior, while the other is obtained by expanding around the theory with very few flavors.

\newsec{Finite temperature}

In this section we will study the finite temperature thermodynamics of systems described by the TDBI action \dbithree. We introduce temperature in the standard way, by replacing the background $AdS_{d+1}$ in which the tachyon field lives with an 
AdS/Schwarzschild black hole which acts as a heat bath. The metric is given by 
\eqn\bhads{  ds^2 = r^2 \left(-F(r) dt^2 + \sum_{i=1}^{d-1}  (dx^i)^2 \right) +{dr^2\over r^2 F(r)};\qquad  F(r) = 1-\left(r_h\over r\right)^d }
where $r_h=\pi T$ is the location of the horizon. $T$ here is the temperature, not to be confused with the tachyon field. To avoid any confusion, in the rest of this section we will use $r_h$ to label the temperature.  

We note in passing that in \KutasovFR\ the temperature was introduced in a slightly different way. There, the background in which the tachyon lived was always $AdS_5\times S^5$, and at finite temperature we replaced it by an $AdS_5$ black hole, which amounts to taking $F(r)=1-(r_h/r)^4$ for all $d$ in \bhads. Presumably, the different ways of introducing temperature should give similar results for the thermodynamics.

The TDBI action (density) at finite temperature is given by
\eqn\dbit{\SS= - \int dr r^{d-1}V(T)  \sqrt{ 1 + F(r) (r\p_r T)^2  },}
and the equation of motion which follows from it is
\eqn\eomtt{   { r^{d-1} \p_T \ln V(T) \over \sqrt{1+F r^2 T'(r)^2} }        = {\p\over\p r} \left( {r^{d+1} F T'\over \sqrt{1+F r^2 T'(r)^2} }  \right).}
As a check, for zero temperature $(r_h=0)$, \eomtt\ reduces to \eomttt. It is useful to change variables from $r$ to
\eqn\defzzz{y=\left(r\over r_h\right)^d,}
in terms of which the action \dbit\ takes the form 
\eqn\dbiz{\SS= - {1\over d} r_h^d\int dy V(T)  \sqrt{ 1 + F(y) (dyT')^2 },  }
where $F(y)=1-{1\over y}$ and $T'=\partial_y T$. Note that in \dbiz, the explicit dependence on $r_h$ has been factored out. The corresponding equation of motion, 
\eqn\eomzz{   {  \p_T \ln V(T) \over \sqrt{1+F(y)(dyT')^2} }        = d^2{\p\over\p y} \left( {y^2 F(y) T'\over \sqrt{1+F(y)(dyT')^2} }  \right), }
is independent of $r_h$ when written in terms of the variable $y$. $y$ ranges between 1 and $(\Lambda/r_h)^d$. We will mostly work in the BKT limit discussed in \KutasovFR\ and the previous sections, in which $1\le y<\infty$. 

Some aspects of the finite temperature analysis are different for the hard wall (in which the potential $V(T)$ vanishes at a finite value of $T=T_{IR}$) and soft wall (infinite $T_{IR}$) models. We will start by discussing the situation for the soft wall case, and then mention the details that are different in the hard wall one. 

We saw in section 3 that at zero temperature, the ground state of soft wall models corresponds to a tachyon configuration of the sort plotted in figure 5. To find the configuration at finite temperature,  we can proceed as follows. Denote by $T_h$ the value of the tachyon field at the horizon,
\eqn\tachhor{T(y=1)=T_h.}
Assuming that $T(y)$ is smooth near $y=1$, the equation of motion \eomzz\ then determines $T'(y=1)$ to be 
\eqn\tprime{T'(y=1)={1\over d^2}\p_T \ln V(T_h).}
Thus, the entire solution of \eomzz\ is determined by specifying $T_h$. 

The value of $T_h$ is determined by the temperature $r_h$ in the following way. As $y\to\infty$ the tachyon field goes to zero, and we can  replace the full equation of motion \eomzz\ by its linearized version, 
\eqn\zzzttt{{\partial\over\partial y}\left[y(1-y)T'\right]-{1\over 4}T=0.}
The general solution of this equation behaves at large $y$ as
\eqn\largetz{T(y)\simeq {1\over\sqrt y}\left(c_1\ln y+c_2\right)} 
where the constants $c_i$ are functions of $T_h$. 

On the other hand, as discussed in \KutasovFR, the UV boundary conditions imply that the asymptotic large $y$ behavior of the finite temperature profile should be the same as at zero temperature. The latter is given by eq. \bktlarger, which can be written in terms of the coordinate $y$ (up to an unimportant overall constant) as 
\eqn\bktlargez{T_0(y)\simeq{1\over\sqrt y}\left({C_1\over d}\ln y+C_2+C_1\ln{r_h\over\mu}\right).}
The ratio of the coefficients of the constant and log terms in \largetz\ and \bktlargez\ must be the same \KutasovFR,
\eqn\toromuro{{1\over d}{c_2(T_h)\over c_1(T_h)}=   {C_2\over C_1} +\ln{r_h \over\mu}. }
Given a potential $V(T)$ and a definition of the scale $\mu$, \toromuro\ gives a relation between $r_h$ and $T_h$. 

A useful observation for the discussion below is that the value of the tachyon at the horizon, $T_h$, goes to zero at a finite value of the temperature, $r_h=r_h^{\rm (crit)}$. To determine  $r_h^{\rm (crit)}$, one can proceed as follows. For small $T_h$, the full solution of \eomzz\ is in the small field regime, and one can determine $T(y)$ by solving \zzzttt. This is the hypergeometric equation with $a=b=1/2$, $c=1$. Its generic solution diverges (logarithmically) as $y\to 1$ and is thus not suitable for our purposes. The solution  that satisfies the boundary conditions $T(y=1)=T_h$ is 
\eqn\sollinear{T(y)={2\over\pi}T_hK(1-y)}
where $2K(x)/\pi$ is the hypergeometric series
\eqn\hyperseries{{2\over\pi}K(x)=1+\sum_{n=1}^\infty \left((2n-1)!!\over n!\right)^2\left(x\over4\right)^n.}
$K(x)$ is known as the complete elliptic integral of the first kind. As $y\to\infty$, it behaves like 
\eqn\largehyper{K(1-y)\simeq {1\over\sqrt y}\left(\half\ln y+2\ln2+\cdots\right).} 
Comparing to \largetz, we see that as $T_h\to 0$, the ratio $c_2/c_1$ approaches the finite limit 
\eqn\limcc{\lim_{T_h\to 0}{c_2(T_h)\over c_1(T_h)}=4\ln2.}
Plugging into \toromuro, we find the critical temperature
\eqn\rhcrit{r_h^{(\rm crit)}=2^{4\over d}\mu\exp\left(-{C_2\over C_1}\right),}
which is given in terms of quantities characterizing the zero temperature problem, namely $\mu$ and $C_i$.  

As $r_h$ approaches $r_h^{(\rm crit)}$, $T_h$ goes to zero. In this limit, the free energy of the non-trivial solution of \eomtt\ approaches that of the trivial solution $T=0$. In general, the difference between the two is given (up to an overall multiplicative constant) by 
\eqn\deltaf{\FF(r_h)=\int_{r_h}^\Lambda dr r^{d-1} \left[V(T) \sqrt{ 1 + F(r) (r\p_r T)^2  }-1\right].}
Here we have returned to the $r$ parametrization, and are working at a large but finite cutoff $\Lambda$, in order not to have to worry about UV divergences. The BKT limit can be taken at the end of the calculation. 

We would like to evaluate the derivative of $\FF$ w.r.t. $r_h$. For this purpose we consider the quantity
\eqn\difffP{\delta\FF=\FF(r_h+\delta r_h)-\FF(r_h).} 
This difference receives contributions from two sources:
\item{(1)} The range of the $r$ integral is different in the two terms. Thus, one contribution to the difference is
\eqn\controne{\delta\FF^{(1)}=-\int_{r_h}^{r_h+\delta r_h}dr r^{d-1}  \left[V(T) \sqrt{ 1 + F(r) (r\p_r T)^2  }-1\right].}
To first order in $\delta r_h$ we can evaluate the integrand of \controne\ at $r=r_h$, and multiply by $\delta r_h$, yielding
\eqn\dfone{\delta\FF^{(1)}=\delta r_hr_h^{d-1}(1-V(T_h))}
where we used the fact that $F(r_h)=0$ and $\partial_r T(r=r_h)$ is finite (see eq. \tprime). 
\item{(2)} When we change $r_h$ by $\delta r_h$, the whole solution $T(r)$ changes to $T(r)+\delta T(r)$. 
If we denote the integrand in \deltaf\ by $\CL(T,T')$, the second contribution to \difffP\ is 
\eqn\contrtwo{\delta\FF^{(2)}=\int_{r_h+\delta r_h}^\Lambda dr\left[ \CL(T+\delta T,T'+\delta T')-\CL(T,T') \right].}
Since $\delta T$ is of order $\delta r_h$, we can replace the lower limit of integration in \controne\ by $r_h$, and write
\eqn\ccctwo{\delta\FF^{(2)}=\int_{r_h}^\Lambda dr\left[ {\delta\CL\over\delta T}\delta T+{\delta\CL\over\delta T'}\delta T' \right]={\delta\CL\over\delta T'}\delta T|_{r_h}^\Lambda. }
In the last step, we integrated the second term by parts and used the equation of motion of $T$. The contribution from $r=\Lambda$ vanishes since $\delta T$ is zero there. The one from $r=r_h$ vanishes as well, since $\delta\CL\over\delta T'$ is proportional to $F(r_h)$, which vanishes. Thus, $\delta\FF^{(2)}=0$. 

\noindent
To summarize, we conclude that 
\eqn\exactder{{\partial\FF\over\partial r_h}=r_h^{d-1}(1-V(T_h)).}
This is a non-trivial equation, since $T_h$ itself is a function of $r_h$, which we have to find first, given the potential $V(T)$. Once we know it, we can solve \exactder\ to find $\FF(r_h)$. A general property we can read off from \exactder\ is that $\FF$ is a monotonically increasing function of $r_h$. In appendix B  we numerically study the finite temperature behavior for some particular soft wall potentials; our results are compatible with \exactder. 

Expanding \exactder\ around $r_h^{\rm (crit)}$ \rhcrit, we find to leading order:
\eqn\smallth{{\partial\FF\over\partial r_h}={d^2\over8} r_h^{d-1}T_h^2+O(T_h^4).}
We can also compute $\FF$ for small $T_h$ directly, by expanding \deltaf\ in a power series in $T$. To determine the form of $\FF$ up to order $T_h^n$, one needs to keep contributions to \deltaf\ up to order $T^n$. The leading contribution is quadratic,
\eqn\deltaquad{\FF_2=\half\int_{r_h}^\Lambda dr r^{d-1} \left[F(r) (rT')^2  +m^2 T^2\right].}
Integrating the first term by parts we get the e.o.m. \zzzttt, so the integral \deltaquad\ reduces to evaluating $r^{d+1}F(r) TT'$ at the boundaries $r=r_h,\Lambda$. Both of these vanish, since $F(r_h)=0$ and $T(\Lambda)=0$. 

Thus,  the leading contribution to $\FF$ is quartic in $T_h$. To calculate it we need to include the quartic contribution to the potential $V(T)$,
\eqn\quarticv{V(T)=1+\half m^2 T^2+{a\over4} T^4+\cdots}
Plugging \quarticv\ into \deltaf, we find
\eqn\deltaquartic{\FF_4=\half\int_{r_h}^\Lambda dr r^{d-1} \left[F(r) (rT')^2  +m^2 T^2+{a\over2} T^4+\half m^2F(rTT')^2-
{1\over4} F^2(rT')^4 \right].}
The solution to the full e.o.m. \eomtt\ can be expanded as $T(r) = T_1(r) + T_3(r) + \dots$ where $T_i(r)$ is of order $T_h^i$. In particular, $T_1$ is given by \sollinear, while $T_3(r)$ is obtained by varying \deltaquartic\ and keeping only terms cubic in $T_h$. 

Using the e.o.m. from \deltaquartic\ and plugging back into $\FF_4$, one finds (in the BKT limit)
\eqn\deltaquarticthree{\FF_4 = \frac 18\int_{r_h}^\infty drr^{d-1}(-6aT^4 + {3d^2\over2}r^2F T^2T'^2+ 3r^4F^2T'^4).
}
Since all terms in \deltaquarticthree\ are quartic in $T$, to leading order in $T_h$ we can replace $T$ by $T_1$ \sollinear. Upon performing the integral, we arrive at
\eqn\deltaquarticfour{ \FF_4 ={1\over d} T_h^4(r_h^{\rm (crit)})^d(c_1 a + d^4 c_2)
}
where
\eqn\cs{
c_1 \simeq -2.847, \qquad c_2 \simeq 0.072.
}
Thus, we see that  the leading behavior of $\FF$ is 
\eqn\quarticff{\FF=cT_h^4+O(T_h^6)}
where
\eqn\cccc{c={1\over d}(r_h^{\rm (crit)})^d(c_1 a + d^4 c_2).}
To compare to \smallth, we need to differentiate \quarticff\ w.r.t. $r_h$. This gives 
\eqn\smallthold{{\partial\FF\over\partial r_h}=4cT_h^3T_h'+\cdots}
where $T_h'$ is the derivative w.r.t. $r_h$. Comparing \smallth\ and \smallthold\ and substituting $r_h=r_h^{\rm (crit)}$, which is okay to leading order in $T_h$, we find 
\eqn\newrh{T_h^2={d^2\over 16c}(r_h^{\rm (crit)})^{d-1}(r_h-r_h^{\rm (crit)})+\cdots}
We are now ready to discuss the finite temperature phase structure of soft wall TDBI models. The phase structure is different for $c<0$ and $c>0$, and we will discuss the two cases separately. 

Consider first the case $c<0$. There are two regions in the phase diagram in which we understand the properties of the non-trivial solution $T(r)$: low temperature (\ie\ small $r_h$), and the vicinity of the critical temperature \rhcrit. As $r_h\to 0$, the value of the tachyon field at the horizon $T_h\to\infty$, and the free energy $\FF$ approaches a finite negative constant. Near $r_h^{\rm (crit)}$, $T_h$ is small. Note that eq. \newrh\ implies that the non-trivial solution only exists for $r_h<r_h^{\rm (crit)}$. The free energy goes to zero as well,
\eqn\leadingf{\FF\sim -(r_h-r_h^{\rm (crit)})^2,}
as can be seen by combining \quarticff\ and \newrh. 

The most natural way to connect the two behaviors is depicted in the left two panels in figure 6. These figures imply that the system exhibits a second order phase transition at $r_h=r_h^{\rm (crit)}$. In the low temperature phase, the solution with lowest free energy has a non-trivial $T(r)$. As $r_h$ approaches the critical value, it smoothly goes to the trivial solution $T=0$, and above the transition temperature, only the trivial solution exists.

\ifig\loc{Qualitative behavior of $r_h(T_h)$ and $\FF(r_h)$ for $c<0$ (left) and $c>0$ (right).}
{\epsfxsize3.3in\epsfbox{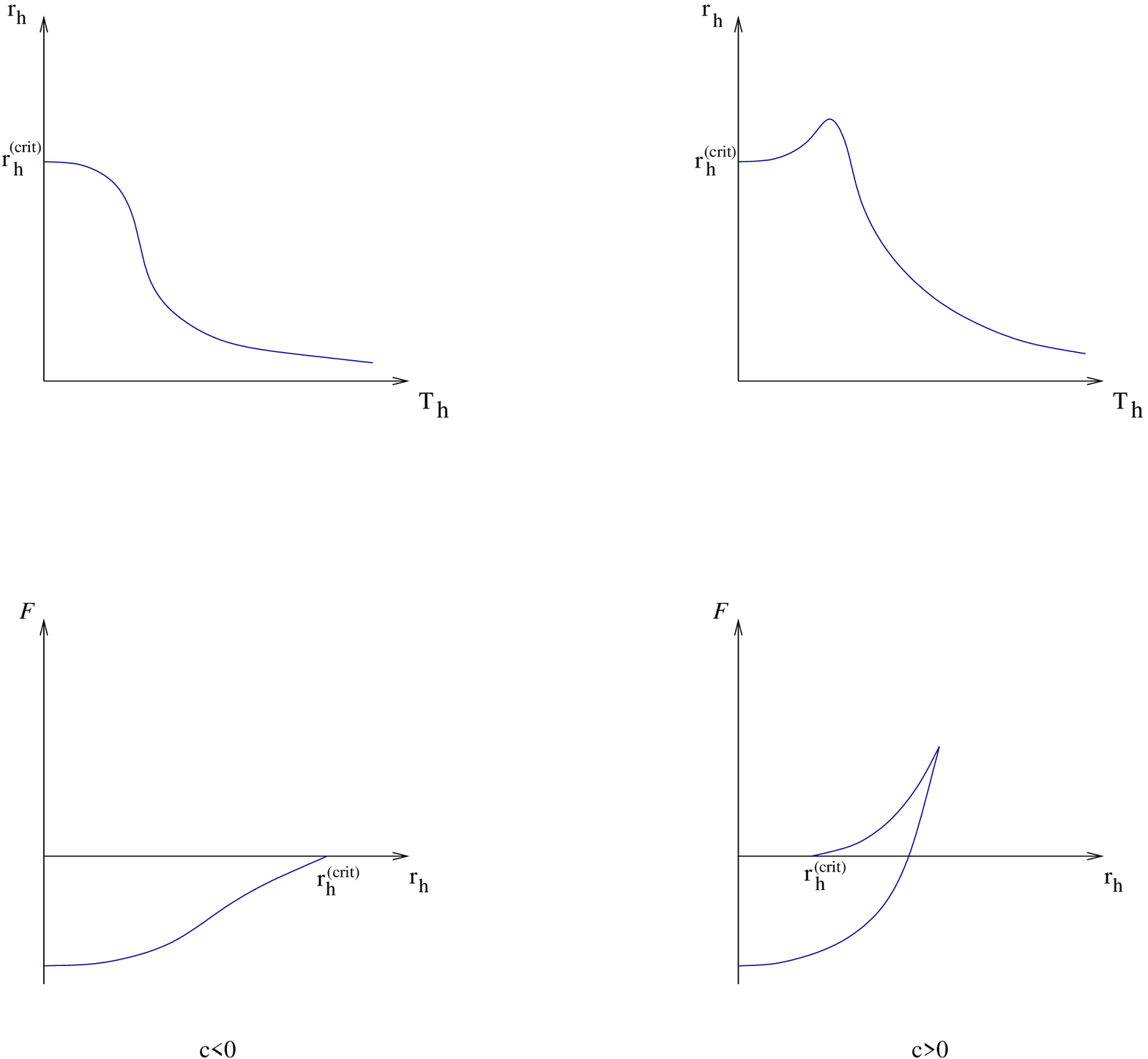}}
\bigskip

For $c>0$ the situation is the following. At low temperatures (small $r_h$) things are as before, but near $r_h=r_h^{\rm (crit)}$ they are different. In particular, \newrh\ implies in this case that there is a non-trivial solution with arbitrary small $T_h$, but it exists only for $r_h>r_h^{\rm (crit)}$. The natural way to connect the two behaviors is depicted in the right panels in figure 6. The upper panel shows $r_h(T_h)$. Starting from $T_h=0$, $r_h$ must initially increase, as indicated by \newrh, but eventually it must turn around and asymptote to zero as $T_h\to\infty$. Assuming that there is a single turning point leads to the upper right plot in figure 6. 

The corresponding behavior of $\FF(r_h)$ is depicted  in the lower right panel of figure 6. At low temperature, $\FF$ is negative and grows with $r_h$. At a maximal value of $r_h$, which can be read off the upper right panel, it connects to another branch of solutions that joins smoothly to the behavior near $T_h=0$ \newrh. In this case, the transition is first order; it happens when the free energy of the low temperature branch of non-trivial solutions approaches that of the trivial, $T=0$ one. It  occurs at the temperature at which the lower curve in the lower right panel in figure 6 intersects the $r_h$ axis. 

The strength of the first order phase transition is controlled by the size of the coefficient $c$ \cccc, which in turn is determined by the coefficient $a$ of the quartic term in the potential \quarticv. As $c$ decreases, the transition becomes less strongly first order,  until eventually $c$ flips sign and the transition becomes second order, as discussed above. If one tunes $a$  such that $c$ vanishes, the leading behavior of $\FF$ near $T_h=0$ is as a higher power of $T_h$ than \quarticff, and one must redo the analysis taking into account higher order terms in $V(T)$. We will leave this analysis to future work.  

The plots of figure 6 were obtained by assuming the simplest possible behavior given the analytic results obtained above. To see that this is indeed what happens, we analyze in appendix B some particular potentials, and verify numerically that the behavior of figure 6 is indeed reproduced, both for positive and negative $c$. 

Our discussion so far has concerned soft wall models. Some aspects of the analysis are the same in the hard wall case. The differential equation  \exactder\ that encodes the dependence of the free energy on temperature, and in particular the monotonicity of this function is still valid,  as is the analysis of the small $T_h$ region, which only relies on the form of the potential \quarticv\ for small $T$.  

On the other hand, the small $r_h$ (low temperature) analysis is different for soft and hard wall models. For the soft wall case,  the solution $T(r)$ ends on the horizon for any finite temperature, and the maximal value of $T(r)$, which we denoted by $T_h$ above, is finite (and thus strictly below the infinite $T_{IR}$). For hard wall models, if the temperature is low enough the vacuum solution $T_0(r)$ does not reach the horizon, and the maximal value of $T_0(r)$ is equal to $T_{IR}$. 

The phase diagram is still expected to be different for $c<0$ and $c>0$. In the former case, it is expected to be given by the left panel in figure 7. For small $r_h$, the non-trivial solution $T(r)$ does not intersect the horizon; its free energy is given by the blue curve. At some value of the temperature, corresponding to the point where the blue and red curves meet, the solution first touches the horizon, and above this temperature it behaves like in the soft wall discussion above. In particular, the value of the tachyon on the horizon $T_h$ is smaller than $T_{IR}$, and goes to zero at  a critical temperature $r_h^{\rm (crit)}$, which corresponds to a second order phase transition. 

\ifig\loc{Qualitative behavior of $r_h(T_h)$ for hard wall models with $c<0$ (left) and $c>0$ (right). The red (blue) curve corresponds to solutions that do (do not) intersect the horizon.}
{\epsfxsize3.3in\epsfbox{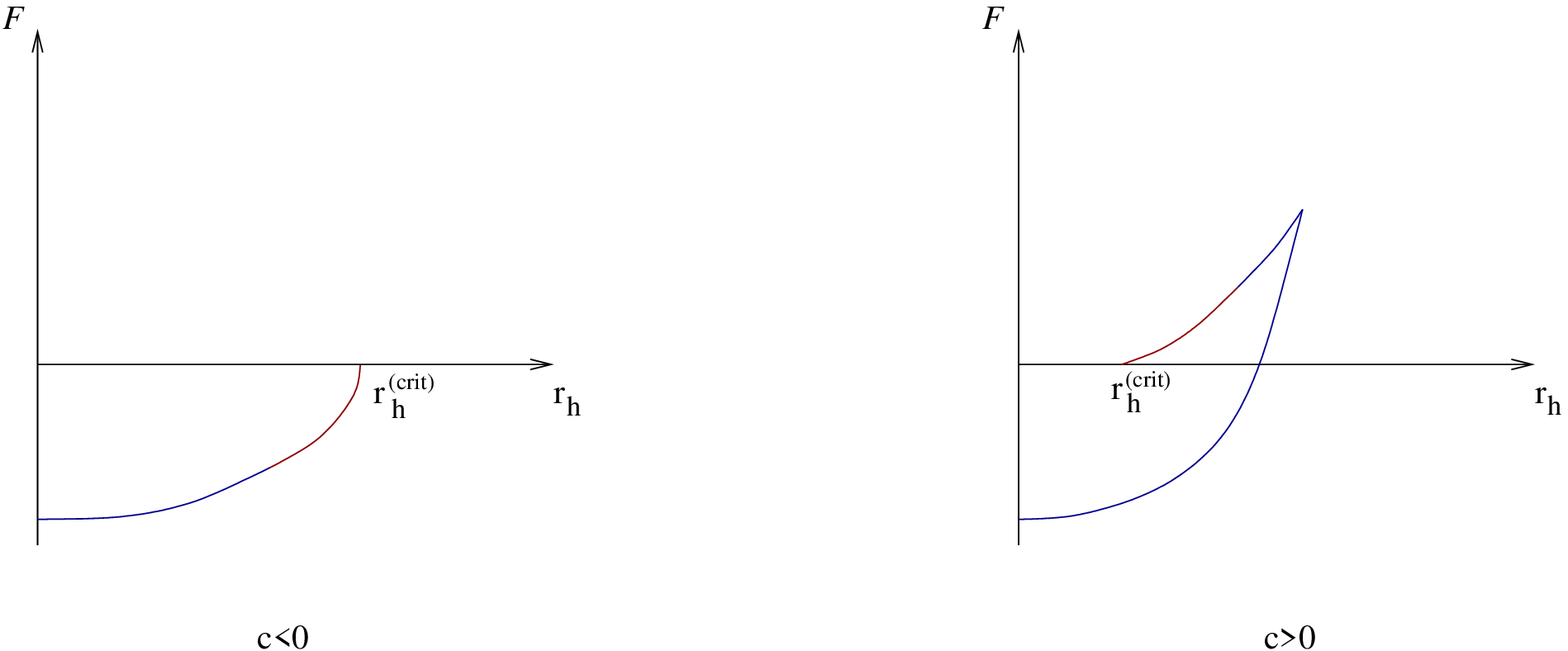}}
\bigskip

For $c>0$ we expect a phase diagram of the sort exhibited in the right panel of figure 7. The blue curve corresponds again to solutions that do not intersect the horizon, while the red curve labels solutions that do. The phase transition is in this case first order; it occurs at the temperature at which the blue line intersects the $r_h$ axis. The situation here is very similar to that analyzed in section 5 of \KutasovFR\ (see figure 7 in that paper), and we refer the reader there for further discussion.  

In summary, for both classes of tachyon potentials, which give rise to hard and  soft wall models, changing the
quartic term in the potential leads to an interpolation between first and the second order phase transitions.
This is in contrast with other holographic models in which the transition is first order, as in models with geometric confinement \WittenZW, probe D-branes \refs{\KruczenskiUQ\AharonyDA\ParnachevDN-\MateosNU} and  hard and soft wall models \HerzogRA. Second order phase transitions in holographic systems have been observed before in holographic superconductors \refs{\HartnollKX}. Our setup is different; it will be interesting to explore possible phenomenological implications.

\newsec{Discussion}

In this paper we studied a class of continuous phase transitions in $d$ dimensional (large $N$) QFT known as conformal phase transitions . We argued that in studying these transitions one can focus on the dynamics of the order parameter, an operator whose dimension $\Delta$ approaches $d/2$ near the transition. Since these transitions often occur at strong coupling, it is natural to use holography to study them. In the bulk description, this corresponds to analyzing the dynamics of a scalar field $T$  in $AdS_{d+1}$, whose mass is close to the BF bound. When the squared mass of $T$ goes below the bound, the theory develops dynamically a small mass gap $\mu$. Interestingly, the universal physics near the transition\foot{I.e. the physics at the scale $\mu$, which goes to zero at the transition and hence can be taken parametrically smaller than all other scales in the system.} is sensitive to the full non-linear Lagrangian of $T$. We took this Lagrangian to have the tachyon DBI form \dbithree, \complextdbi, and analyzed the dynamics near the transition for different potentials $V$. This can be viewed as a bottom-up description, in the spirit of AdS/QCD and AdS/CMT studies. 

We found that the dynamics of this class of models is similar to that of hard and soft wall models of AdS/QCD \refs{\ErlichQH,\KarchPV}, with a dynamically generated wall. Hard wall models are obtained when the potential $V(T)$ vanishes at a finite value of $T$, $T_{IR}$, while soft wall ones correspond to potentials with $T_{IR}\to\infty$. For hard wall models, the spectrum of small excitations (mesons) near the transition was found to take the standard Kaluza-Klein form \mmmnnn\ at large excitation level, while for soft wall ones  we exhibited potentials with non-trivial high energy spectra \thooftas, \asmn. We also studied the finite temperature thermodynamics of these models and found that they exhibit first or second order phase transitions, depending on the potential $V(T)$. Interestingly, the high energy spectrum of mesons and the type of finite temperature phase transition exhibited by the model are sensitive to different features of the potential. The spectrum is sensitive to the behavior of $V(T)$ near the bottom (see figure 4), while the finite temperature phase transition is sensitive to the behavior near the maximum at the origin of field space. 

In this section we would like to comment on some  aspects of our construction and possible generalizations. One natural question is why it is sensible to focus on the dynamics of the bulk tachyon field $T$ dual to the boundary operator $\OO$ whose dimension approaches $d/2$ near the transition, and neglect the backreaction of other fields on it.  The reason is that near the transition one can think of the boundary theory as a CFT deformed by the ``double trace'' operator $\OO^2$. Correlation functions in the deformed theory, such as the two point function $\langle\langle\OO(x)\OO(0)\rangle\rangle$, are given by expressions like
\eqn\confpert{\langle\langle\OO(x)\OO(0)\rangle\rangle=\langle\OO(x)\OO(0)\exp\left[-\lambda\int d^dz\OO^2(z)\right]\rangle}
where the expectation value on the right hand side is computed in the CFT. Thus,  the two point function \confpert\ can be calculated in conformal perturbation theory in terms of (integrated) correlation functions involving multiple $\OO$ operators in the CFT. 

The dual statement in the bulk theory is that in order to calculate properties of the massive theory, all we need to know is the non-linear Lagrangian for the order parameter $T(x^M)$. To compute correlation functions of other operators in the boundary theory, such as the global symmetry currents discussed above, we similarly need to know the non-linear terms in the Lagrangian involving the corresponding bulk fields and an arbitrary number of $T$'s, which is precisely the type of information encoded in actions such as \complextdbi. Of course, the above discussion does not imply that the dynamics of $T$ should necessarily be governed by the TDBI action with any potential. The assumption that it is, is the bottom-up aspect of our analysis. It would be nice to provide further justification for this assumption or to improve on it.

It is important to stress that our construction is quite different from what is known in the study of holographic systems as the probe approximation. In the context of QCD, this approximation is valid in the 't Hooft limit, in which the number of colors $N$ goes to infinity while the number of flavors $F$ is kept fixed. In this limit, the leading contribution to the bulk action involves the metric and other closed string fields, and takes the form of Einstein gravity with various corrections.  Confinement should be a property of this action, and manifest itself via the generation of a hard or soft wall in the geometry.

The first subleading term in $F/N$ involves fields associated with the quarks, such as the tachyon which is dual to the meson operator, the gauge fields dual to the global currents etc. Their action is usually taken to be of the TDBI form \complextdbi\ with corrections (see \eg\ \CaseroAE,\BigazziMD). Although this term is subleading in $F/N$, it is the leading contribution to the dynamics of fields in the meson sector, and clearly needs to be kept when studying meson dynamics. 

As $F/N$ increases, the probe approximation becomes worse and worse, until eventually when $F\sim N$, it breaks down. In general, in this region of parameter space we have no reliable tools for studying the dynamics for energies of order $\Lambda_{QCD}$ and below, where the theory is strongly coupled. The main point of our construction is that as $F\to F_c$ (\ie\ near the transition from conformal to confining behavior), we again expect to be able to provide a good bulk description of the dynamics. However, the small parameter is now not $F/N$, as in the probe approximation, but the ratio of scales $\mu/\Lambda_{QCD}$, where $\mu$ is the dynamically generated meson mass scale, and $\Lambda_{QCD}$ is the crossover scale between the free UV and interacting IR CFTs. 

As we argued, near the transition we can describe the dynamics by writing an action for the bulk field $T$ dual to the meson operator. However, this action plays a very different role from that which appears as the first subleading contribution to the bulk action in the probe region $F\ll N$. In particular, we should not add to the bulk Lagrangian an Einstein term for the metric and dilaton fields, which would drastically modify the dynamics of the tachyon. The interpretation of this modification in the boundary theory would be that in addition to the operator $\OO^2$, the Lagrangian includes non-trivial couplings of other marginal or relevant operators, which is not expected to be the case near the transition. 

Another interesting issue concerns the chiral symmetry of the complex TDBI action \complextdbi. In the massive phase, chiral symmetry is broken by the non-trivial condensate of the tachyon field $T$. This condensate goes to zero at large $r$ (or small $z$); therefore, the symmetry is restored there. In the boundary theory, the restoration is reflected in the short distance structure of off-shell Green functions of the currents dual to the gauge fields  $A^{(L,R)}$. 

Naively, one might expect the restoration to also be visible in the highly excited spectrum of mesons. However, we found by an explicit calculation that for potentials of the form \inftir\ (at large $T$), the vector and axial vector spectra remain distinct at arbitrarily large excitation levels.\foot{A similar result was found in \CaseroAE. However, we also found an example \ttlltt\ where the highly excited spectra agree, which was not the case there.} This is not surprising given that the highly excited spectrum depends on the large $z$ (or small $r$) behavior of the vacuum solution $T_0(r)$. This region is usually thought of as the IR region in terms of the usual correspondence between scale in the field theory and position in the radial direction in $AdS$, and in it the symmetry breaking effects due to the non-zero $T_0(r)$ are not suppressed.

At first sight it is surprising that the masses of highly excited mesons, which are very heavy, are sensitive to IR physics. 
In fact, this is a phenomenon familiar from black hole physics and holography, known as the UV/IR correspondence. The TDBI analysis incorporates the fact that highly excited mesons are physically large (in $\IR^d$), hence their properties are sensitive to the long distance behavior of the theory. In particular, there is no contradiction between the fact that chiral symmetry is restored at short distances in off-shell Green functions, and its lack of restoration in the highly excited meson spectrum (see \refs{\ShifmanXN,\IatrakisJB} and references therein for further discussion). In fact, the TDBI system provides a nice laboratory for studying the relation between the two in a large class of models. 

One of the important applications of strongly coupled models of the sort we studied is to electroweak symmetry breaking (technicolor). In this context, the system analyzed in this paper can be viewed as the symmetry breaking sector of walking technicolor. We will leave a more detailed investigation of the resulting phenomenology to future work. One aspect that we would like to mention is the techni-dilaton, a pseudo Nambu-Goldstone boson of broken conformal symmetry, which may play an important role in current collider experiments (see \eg\ \CampbellIW\ for a recent discussion). 

In our model, the dilaton is the lightest $\sigma$-meson, \ie\ the lightest normalizable eigenstate of the bound state potential  \schrodinger, \effpot. A natural question is whether this state is expected to be anomalously light, compared to other scalar and vector mesons. We have not studied this problem in our paper, but qualitatively  expect this to be the case, for the following reason. 

As discussed in section 3, the effective potential for $\sigma$-mesons \effpot\ behaves at small $z$ like \veffir, and as explained there, if this was the exact behavior of the potential, the spectrum would contain tachyons with mass squared of order $-\mu^2$. The potential \effpot\ actually deviates from \veffir\ at $z\sim 1/\mu$ and larger, which raises the mass squared of the lowest lying $\sigma$-meson by an amount proportional to $\mu^2$. Thus, it is natural to expect the mass of the lowest lying meson to be of order $\MM^2=\alpha\mu^2$ with $\alpha$ smaller than the typical separation between states. In \KutasovFR\ we analyzed a particular (hard wall) model and found that this is indeed the case.  

Note that $\alpha$ cannot be negative, since that would imply that the configuration $T=T_0(r)$ is unstable, contrary to assumption. For generic potential $V(T)$ we would expect it to be non-zero. We do not know whether it is possible to tune the potential such that $\alpha$ vanishes, giving a massless techni-dilaton. Since the breaking of conformal symmetry in our model is explicit, we would expect this not to be the case. It would be interesting to see how small one can make $\alpha$ by tuning the potential.

Another natural question is whether one can study meson dynamics in QCD using our approach. Our model is rather different from other holographic QCD models, primarily since the latter attempts to describe the physics in the limit $N\to\infty$, $F$ fixed and then continue to the physical regime, while our approach expands around the boundary between confining and conformal behavior $F=F_c$. The two expansions can potentially provide complementary information about the physics
of mesons.

In Section 6, we studied the dynamics of our model at finite temperature. We found that the phase diagram of the system can exhibit either  first or second order phase transitions, depending on the value of the quartic term in the tachyon potential. Recently there has been a lot of discussion of possible applications of holography to condensed matter physics (see e.g. \refs{\HerzogXV\McGreevyXE\HorowitzGK\HartnollFN-\SachdevWG} for reviews). Second order phase transitions might be interesting in this context, since near such transitions one may hope to extract universal quantities, such as critical exponents. For example, from \leadingf\ one can infer the behavior of the heat capacity near the phase transition: it experiences a jump from a constant non-zero value to zero, which is the standard mean field behavior.
One can tune the tachyon potential  such that the critical exponent  $\alpha$ associated with the specific heat obtains a non mean field value. We will leave a more detailed discussion of this to future work. 

Another interesting feature of our model is its low temperature behavior. In the usual hard and soft wall AdS/QCD models, the low temperature phase is obtained from the zero temperature geometry  by compactifying  Euclidean time. The black hole phase is thermodynamically favored only above a critical temperatures (see \eg\  \HerzogRA). This implies that there is no dissipation at small temperatures, at least in the classical approximation to the bulk theory. In contrast, in our soft wall models, the black hole phase describes the thermodynamics at arbitrarily low temperature, \ie\  the finite temperature vacuum solution $T_0(r)$ extends all the way to the horizon at $r=r_h$. Hence, the effective metric for fluctuations has a horizon, and the system exhibits dissipation at arbitrarily low temperature. A detailed investigation of this will also be left for future work.

\bigskip

\noindent{\bf Acknowledgements}: We thank O. Aharony, E. Kiritsis, M. Kulaxizi,  E. Martinec  and J. Sonnenschein for discussions. This work is supported in part by DOE grant DE-FG02-90ER40560, the BSF -- American-Israel Bi-National Science Foundation and a VIDI innovative research grant from the Netherlands Organisation for Scientific Research (NWO). JL is supported in part by an NSF Graduate Research Fellowship. DK and AP thank the organizers of the 6th Crete Regional Meeting in String Theory for hospitality during part of this work. DK thanks the Weizmann Institute for hospitality during part of the work.

\appendix{A}{Meson spectrum in the axial sector}

In this appendix, we discuss the meson spectrum in the axial sector. Recall that the action for the complex tachyon DBI \complextdbi\ is 
\eqn\complextdbitwo{\SS =-  \int d^{d+1}xV(T^\dagger T)\left(\sqrt{-G^{(L)}}+\sqrt{-G^{(R)}}\right)}
where 
\eqn\openmetrictwo{
G_{MN}^{(L)} = g_{MN} + \frac 12 D_{(M}T^\dagger D_{N)}T+F_{MN}^{(L)}}
and similarly for $L\leftrightarrow R$. The complex tachyon, vector, and axial vector gauge fields are defined to be 
\eqn\vaxialtwo{T = \tau \exp (i\theta), \qquad V=A^{(L)}+A^{(R)}, \qquad A=A^{(L)}-A^{(R)}.}
Expanding to quadratic order, the action for the axial sector is
\eqn\saxial{ S = -\int d^{d+1}x\sqrt{G} V(\tau)\left(\frac 18 G^{MM'}G^{NN'}F^{(A)}_{MN}F^{(A)}_{M'N'} + G^{MN}\tau^2(\partial_M\theta + A_M)(\partial_N\theta+A_N)  \right).}
It is convenient to change to the $z$ coordinate following \cov, in terms of which the action reads
\eqn\saxialtwo{S = -\int d^{d+1}xr(z)^{d+1} V(\tau)\left(\frac 18 r(z)^{-4}F_{(A)}^{\alpha\beta}F^{(A)}_{\alpha\beta} + r(z)^{-2}\tau^2(\partial^\alpha\theta + A^\alpha)(\partial_\alpha\theta+A_\alpha)  \right),}
where indices are raised and lowered with the flat metric $\eta_{\alpha\beta}$. This Lagrangian gives rise to axial vector and pseudoscalar mesons; we will now analyze them in turn, starting with the vectors. 

For the purpose of this discussion we can set $\theta=0$, pick the gauge $A_z = 0$ and parametrize the remaining components as $A_a(z,x) = A_a(x)v(z)$. The transversality condition takes the form $\partial^aA_a = 0$. The equation of motion following from the action \saxialtwo\ is
\eqn\vectoreom{\frac 14\partial_z(V(\tau)r(z)^{d-3}\partial_zv(z)) + \frac 14 m_n^2V(\tau)r(z)^{d-3}v(z) - V(\tau)r(z)^{d-1}\tau^2v(z) = 0. 
}
We make the change of variable $\psi(z) = e^{-B/2}v(z)$ where
\eqn\defbz{B(z) = -\ln V(\tau) - (d-3)\ln r(z).}
Then \vectoreom\ takes the Schr\"odinger form with
\eqn\axialpotential{V_{\rm{eff}} = \frac 1 4 (B')^2 - \frac 12 B'' +  4r(z)^2\tau(r(z))^2. }
Plugging $A_a(z,x)$ into the bulk action \saxial, we find a tower of axial vectors
\eqn\fourdaction{S = -\int d^dx \left(\frac 14 F_{ab}^{(A)}F^{ab}_{(A)} + m_n^2A_a A^a\right)}
subject to the following normalization conditions:
\eqn\normalization{\eqalign{
\int dz V(\tau) r(z)^{d-3}v(z)^2 &= \int dz \psi(z)^2 \simeq 1 \cr
\int dz V(\tau) \left(r(z)^{d-3}(v')^2 + 4 r(z)^{d-1}\tau^2v(z)^2   \right)  &= \int dz\left[\left(\frac{B'}{2}\psi + \psi'\right)^2 + 4r(z)^2\tau^2\psi^2\right]  
= {m^2_n}}}
where we have omitted unimportant overall constants. In the second line, we integrated by parts and used the Schr\"odinger equation. Since the norm is the standard Schr\"odinger one, the spectrum is obtained by solving the bound state problem for a particle moving in the potential \axialpotential. 

Next, we turn to the pseudoscalar modes, and show that the spectrum always contains a massless pion. Parametrizing $A_a(z,x) = \varphi(z)\partial_a\pi(x), \theta(z,x) = \theta(z)\pi(x)$, the equations of motion of \saxialtwo\ are
\eqn\pioneom{\eqalign{
\frac 14\partial_z(V(\tau)r(z)^{d-3}\partial_z\varphi) - V(\tau)r(z)^{d-1}\tau^2(\theta + \varphi) &= 0 \cr
\tau^2\partial_z\theta + \frac 14 m^2 r(z)^{-2}\partial_z\varphi &= 0.
}}
The normalization conditions for the kinetic and mass terms of the field $\pi$ are
\eqn\pionnorm{\eqalign{
\int dz \left(\frac 14 V(\tau)r(z)^{d-3}(\varphi')^2 + V(\tau)r(z)^{d-1}\tau^2(\theta + \varphi)^2\right)  &= 1 \cr
\int dz V(\tau)r(z)^{d-1}\tau^2(\partial_z\theta)^2 &= m_n^2 . 
}}
When $m^2 = 0$, the second equation in \pioneom\ implies that $\theta(z)$ is a constant, which can be set to zero by a gauge transformation. At small $z$ (in the UV), the remaining equation of motion reduces to
\eqn\pionuv{
\frac 14 \partial_z(z^{3-d}\partial_z\varphi) - z(\log z)^2\varphi = 0.
}
The second term is presumably subleading to the first and we drop it. Then the two independent solutions are $\varphi \sim z^{d-2}, c$. Plugging into \pionnorm, it is easy to check that the kinetic normalization condition does not diverge for either solution at small $z$, as long as $d>2$. Since both UV solutions are normalizable, the massless mode exists as long as there is one normalizable solution in the IR. 

Let us confirm that this is true for the hard-wall and soft-wall models discussed in sections 4 and 5. For the hard wall model \irpoten\ at large $z$, $r \sim \mu$ and $\tau \sim \tau_{IR}$ are both finite and have no effect on the divergence structure, so we ignore them here. The eom (up to constants) is
\eqn\hwir{
\partial_z((z_{IR}-z)^\alpha\partial_z\varphi) - (z_{IR}-z)^\alpha\varphi = 0.
}
We can neglect the second term (which turns out to be self-consistent). Then the solutions are $\varphi \sim (z_{IR}-z)^{1-\alpha}, c$. Since we are concerned with $\alpha>1$, there indeed is one normalizable and one non-normalizable solution. 

For the soft wall case, we first consider the potential \inftir, $V(T) \sim e^{-\frac 12 \beta T^2}$. We assume that $\beta >d$ because the alternative gives rise to a continuum of scattering states and is uninteresting for our purposes. The equation of motion in the IR is
\eqn\exppotirtwo{
\partial_z\left(\exp(-\frac 12\beta z^{\frac{2\beta}{\beta+d}})\partial_z\varphi \right) = z^{\frac{2(\beta-d)}{\beta+d}}\exp\left(-\frac 12\beta z^{\frac{2\beta}{\beta+d}}\right)\varphi.
}
With the ansatz $\varphi(z) = f(z)\exp(\frac 12\beta z^{\frac{2\beta}{\beta+d}})$, the eom reduces to
\eqn\eomf{
f'' + \frac{\beta^2}{\beta+d}z^{\frac{\beta-d}{\beta+d}} f'= z^{\frac{2(\beta-d)}{\beta+d}} f 
}
(where again, we take a large $z$ limit at every step). Now, we make the ansatz $f = \exp(a z^{\frac{2\beta}{\beta+d}}$). This yields a quadratic equation for the coefficient $a$:
\eqn\aeqn{ 4a^2\beta^2 + 2a\beta^3 - (\beta+d)^2 = 0,
}
which has two solutions, one with $a>0$ and one with $a < -\beta/2$. One can readily check that the latter is normalizable. The model which gives rise to linear confinement \ttlltt\ can be checked using the same steps and behaves similarly.

Finally, we examine the spectrum of massive pseudoscalar modes. If $m^2$ is not zero, we can solve the first equation in \pioneom\  for $\theta$  and substitute the result into the second one. This leads to
\eqn\eommassivepi{
\partial_z\left(\frac{1}{V(\tau)r^{d-1}\tau^2}\partial_z\Phi \right) - \frac{4}{V(\tau)r^{d-3}}\Phi + \frac{m^2}{V(\tau)r^{d-1}\tau^2} = 0
}
where $\Phi = V(\tau)r^{d-3}\partial_z\varphi$. We make the additional change of variable $\Psi(z) = e^{-B/2}\Phi(z)$ where
\eqn\pseudoscalarb{
B(z) = \ln V(\tau)r(z)^{d-1}\tau(r(z))^2.
}
Then the eom \eommassivepi\ takes the Schr\"odinger form with the effective potential
\eqn\pscalarpot{
V_{\rm{eff}} = \frac 14 (B')^2 - \frac 12 B'' + 4r(z)^2\tau(r(z))^2.
}
However, the normalization condition from the kinetic term is not the usual Schr\"odinger norm. Instead, it is
\eqn\pseudonorm{
\int dz \frac{1}{4V(\tau)r^{d-3}}\Phi^2 + \frac{1}{16V(\tau)r^{d-1}\tau^2}(\partial_z\Phi)^2 = 1.
}

\appendix{B}{Thermodynamics of a particular TDBI model}

In this appendix we  analyze the thermodynamics of a particular class of soft wall potentials in $d=4$, as a check on the general discussion of section 6. These potentials are given by
\eqn\potvar{   V(T) = \left(  {3\over4}+{1\over4}{ 1-A T^4\over 1+A T^4} \right) \exp\left( - 2 T^2 \right)   }
The quadratic term in the potential is tuned to the BKT limit for $d=4$, \ie  $V(T)=1-2T^2+\dots$ The coefficient $A$ controls the value of the quartic term in the potential.  Comparing to \quarticv, we see that $a=8-2A$. 

Our goal is to analyze the thermodynamics for two values of $A$ that give rise to positive and negative $c$ in \cccc. For $d=4$ the transition between the two regimes occurs at $a\simeq 6.5$. Thus, we will analyze two cases: 
\item{(1)} $A=0$, $a=8$: \cccc\ gives $c<0$ in this case, so we expect a second order phase transition.
\item{(2)} $A=5$, $a=-2$: $c>0$, first order phase transition.  

The numerical results for $A=0$ are shown in figures 8, 9. We see that they are in agreement with the discussion of section 6, and in particular the two left panels in figure 6. 

The numerical results for $A=5$ are exhibited in figure 10, 11. They agree with those in the two right panels of figure 6, and imply that the transition is first order in this case.

\ifig\loc{$r_h/\mu$ as a function of $T_h$ for $A=0$}
{\epsfxsize3.1in\epsfbox{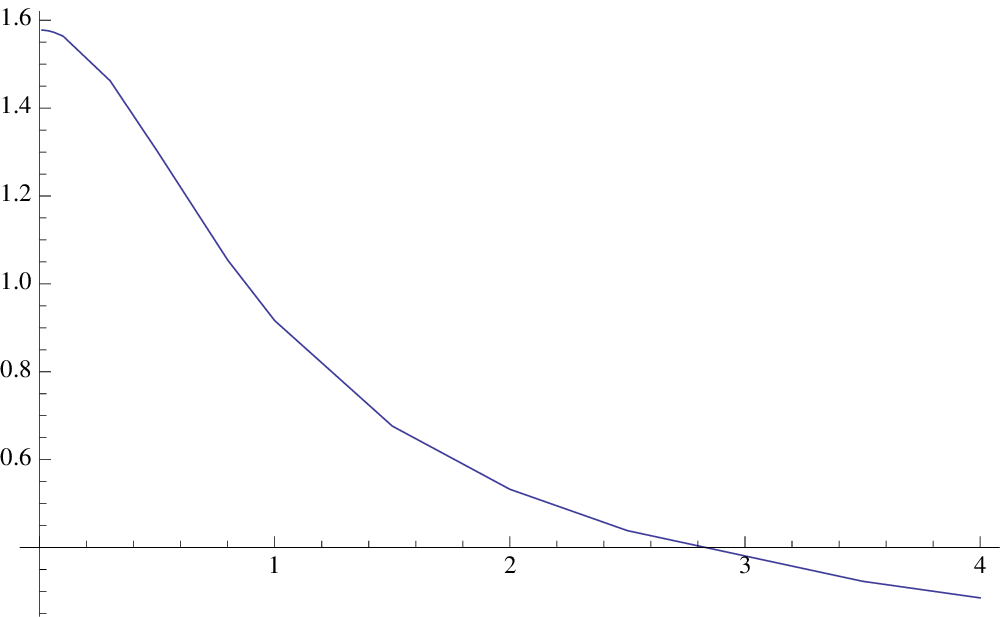}}
\bigskip
\ifig\loc{$\FF/\mu^4$ as a function of $r_h/\mu$ for $A=0$}
{\epsfxsize3.1in\epsfbox{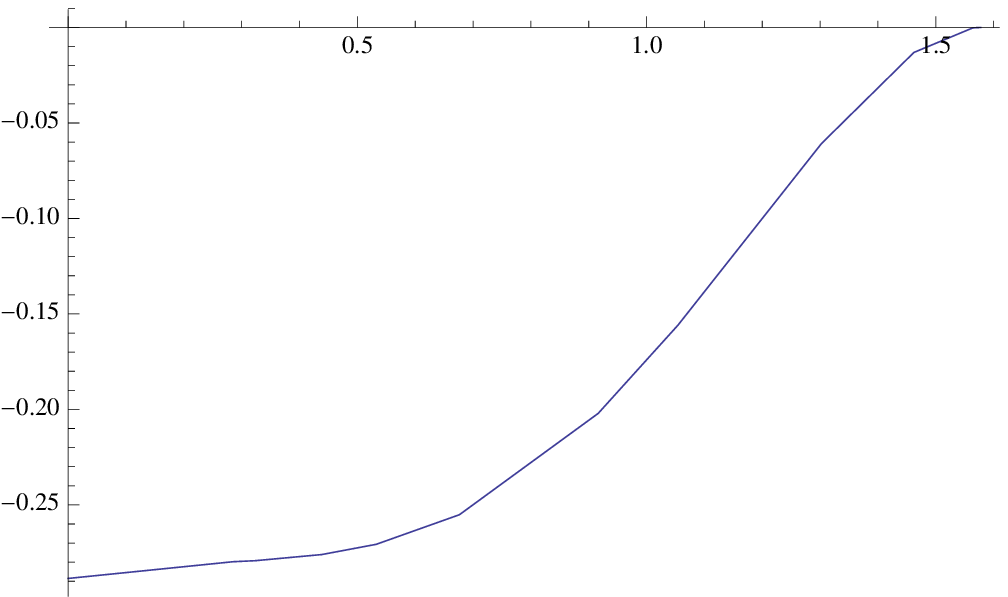}}
\bigskip

\ifig\loc{$r_h/\mu$ as a function of $T_h$ for $A=5$}
{\epsfxsize3.1in\epsfbox{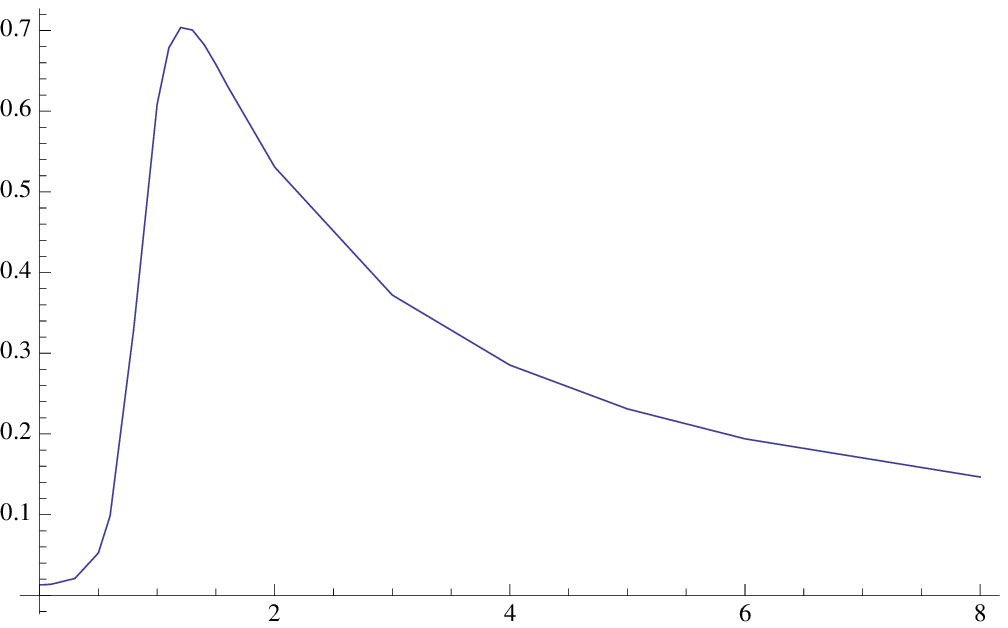}}
\bigskip
\ifig\loc{$\FF/\mu^4$ as a function of $r_h/\mu$ for $A=5$}
{\epsfxsize3.1in\epsfbox{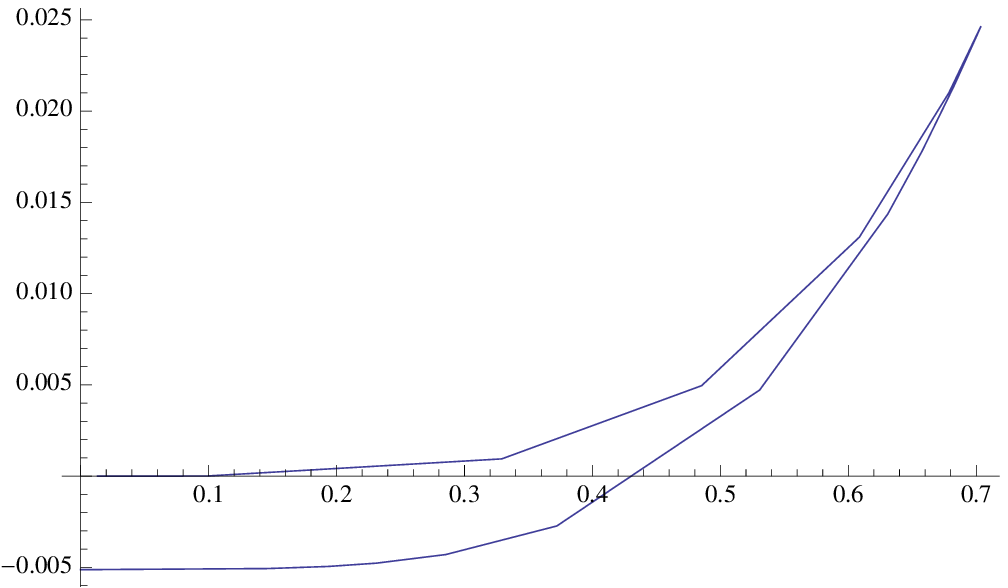}}

\listrefs

\bye